%% file: main.tex
\documentclass[3p,nopreprintline]{elsarticle}
\usepackage{amssymb}
\usepackage{amsmath}

\usepackage[acronym]{glossaries}
\glsdisablehyper
\newacronym{fem}{FEM}{finite element method}
\newacronym{dem}{DEM}{deep energy method}
\newacronym{sem}{SEM}{shallow energy method}
\newacronym{bvp}{BVP}{boundary value problem}
\newacronym{lbfgs}{L-BFGS}{limited-memory Broyden-Fletcher-Goldfarb-Shanno}
\newacronym{pinn}{PINN}{physics-informed neural network}
\newacronym{pde}{PDE}{partial differential equation}
\newacronym{rff}{RFF}{random Fourier feature}
\newacronym{gpu}{GPU}{graphics processing unit}
\newacronym{pam}{PAM}{polyacrylamide}
\newacronym{mlp}{MLP}{multi-layer perceptron}
\newacronym{kan}{KAN}{Kolmogorov-Arnold network}
\newacronym{rbfnn}{RBFNN}{radial-basis-function neural network}
\newacronym{rbf}{RBF}{radial basis function}
\newacronym{vpinn}{vPINN}{variational physics-informed neural network}
\newacronym{gnn}{GNN}{graph neural network}
\newacronym{pa66}{PA66}{polyamide 66}
\newacronym{cfrp}{CFRP}{carbon fiber reinforced polymer}
\newacronym{nem}{NEM}{neural energy method}
\newacronym{cnem}{C-NEM}{collocation-based neural energy method}
\newacronym{mnem}{M-NEM}{mesh-based neural energy method}
\usepackage{upgreek}
\usepackage{booktabs}
\usepackage{tikz}
\usetikzlibrary{arrows.meta, calc, patterns, external}
\usepackage{caption} 
\usepackage[colorlinks=true]{hyperref}
\hypersetup{
    linkcolor=blue,          
    citecolor=blue,          
    urlcolor=blue,           
    pdftitle={A mesh-based neural energy method for the simulation of heterogeneous composites}
}
\usepackage{ulem}
\usepackage{xcolor}
\makeatletter
\newcommand\hl{%
  \bgroup
  \UL@protected\def\sout{\bgroup \ULdepth =-.8ex \ULset}%
  \markoverwith{\textcolor{yellow}{\rule[-.5ex]{.1pt}{2.5ex}}}%
  \ULon}
\makeatother

\newcommand{\B}{\boldsymbol}

\journal{}

\begin{document}

\begin{frontmatter}

\title{A mesh-based neural energy method for the simulation of heterogeneous composites}

\author[aff1]{Pius J. M. Wichmann\corref{cor1}}
\ead{p.wichmann@campus.tu-berlin.de}
\author[aff1]{Stefan Hildebrand}
\cortext[cor1]{Corresponding author}
\author[aff1]{Sandra Klinge}
\affiliation[aff1]{organization={Department of Structural and Computational Mechanics, Technische Universität Berlin},
            addressline={Straße des 17. Juni 135}, 
            city={Berlin},
            postcode={10623}, 
            state={Berlin},
            country={Germany}}

\begin{abstract}
Modeling heterogeneous materials remains a challenge for physics-informed neural networks such as the deep energy method (DEM). The DEM and its variants, here collectively referred to as the neural energy method (NEM), offer a differentiable variational framework. However, their conventional collocation-based implementation (C-NEM) often suffers from physically inadmissible displacement oscillations, integration errors, and high computational costs from automatic differentiation. This work introduces the mesh-based neural energy method (M-NEM), extending the NEM through a mesh-based discretization of the displacement field. By interpolating nodal displacements via shape functions, the M-NEM imposes a kinematic constraint that suppresses oscillations. Furthermore, the method replaces automatic differentiation with algebraic shape function derivatives for strain computation and employs high-order Gaussian quadrature for accurate energy integration. On a directly comparable benchmark problem, the M-NEM reduces stress errors by up to three orders of magnitude relative to the C-NEM while being one to two orders of magnitude faster. On two further benchmarks involving extreme stiffness contrasts, only the M-NEM converges. A comparative study of neural architectures reveals that radial basis function neural networks (RBFNNs) yield optimal performance within the M-NEM, resolving sharp gradients at material interfaces with higher accuracy than multi-layer perceptrons (MLPs) with random Fourier feature (RFF) mapping and faster convergence than Kolmogorov-Arnold networks (KANs).
\end{abstract}

\begin{keyword}
Deep energy method \sep Heterogeneous composites \sep Radial basis function neural networks \sep Physics-informed neural networks \sep Solid body mechanics

\end{keyword}

\end{frontmatter}


\section{Introduction}
\label{sec:introduction}
Heterogeneous composites are fundamental to both engineering and nature, ranging from biological tissues to fiber-reinforced industrial structures. The simulation of these materials is numerically challenging due to the discontinuities and stress concentrations caused by substantial contrasts in stiffness~\cite{klinge2012}. 

In recent years, machine learning approaches for solid mechanics simulations have gained increasing traction~\cite{kollmannsberger2025,herrmann2024}. Most notably, \glspl{pinn}~\cite{raissi2019, karniadakis2021physics} stand out for their potential to efficiently solve geometrically complex, high-dimensional or inverse problems~\cite{sanjalawe2026}, demonstrating remarkable versatility in tasks such as parameter identification, material characterization, and design optimization~\cite{goswami2020,he2023topology,hildebrand2025}.
In this approach, the solution is approximated by a neural network and the loss function is derived from physical laws, typically expressed as the residual of the governing \glspl{pde} and evaluated on a set of collocation points. To reduce the continuity requirements imposed by strong-form \glspl{pde}, \glspl{vpinn} have been proposed~\cite{kharazmi2019, berronne2022, radin2023, anandh2025}. In this framework, samples of test functions are evaluated to enforce compliance with the weak form of the \gls{pde}. In elastostatics, the principle of minimum potential energy is employed directly by the \gls{dem}~\cite{samaniego2020, nguyenthanh2020}, eliminating the need to evaluate test functions. In contrast to strong-form \glspl{pinn}, the \gls{dem} requires only first-order derivatives of the displacement field.

The structure of the \gls{dem} is largely modular, allowing for the independent exchange of core components~\cite{wang2025xdem}. Specifically, the standard \gls{mlp} can be enhanced by input encoding via \gls{rff} mapping~\cite{chadha2023} or replaced by other network architectures, such as \glspl{rbfnn} in the \gls{sem}~\cite{liang2022}, \glspl{kan}~\cite{liu2025,wang2025kan}, and \glspl{gnn}~\cite{gao2022, heGNN2023, brumand2025}. Alternatively, nodal values in a finite element interpolation are optimized directly~\cite{thombre2025}, or their shape functions constructed by a neural network and optimized together with the mesh coordinates in the HiDeNN-FEM~\cite{zhang2021}. To unify the terminology, this broad group of energy minimization approaches is referred to as \gls{nem} in the following. The evaluation of the energy functional is conventionally performed on collocation points, and is referred to hereafter as the \gls{cnem}. The numerical integration scheme used to compute the total potential energy has been shown to be crucial for the accuracy of the \gls{nem}~\cite{nguyenthanh2020}. Various approaches have been implemented, including Monte Carlo sampling~\cite{samaniego2020}, Newton-Cotes formulas~\cite{nguyenthanh2020}, domain-wide Gaussian quadrature~\cite{nguyentanh2021}, and quadrature based on Delaunay triangulation~\cite{fuhg2022}. 

Despite these advancements, the application of the \gls{cnem} to heterogeneous composites faces several challenges. (i) Complex structures require a substantial number of collocation points to limit physically inadmissible displacement oscillations, a phenomenon here referred to as geometric overfitting. (ii) Computing the deformation gradient via automatic differentiation increases computational overhead and memory consumption. (iii) In heterogeneous materials where the strain energy density varies abruptly across inclusions, stochastic or low-order integration schemes introduce significant aliasing errors. Finally, a fourth, independent challenge arises from the network architecture itself: conventional \glspl{mlp} suffer from spectral bias, prioritizing the learning of low-frequency features while struggling to capture the sharp gradients present at material interfaces~\cite{tancik2020}.

To address these issues, the current work proposes the \gls{mnem}, extending the \gls{nem} through a mesh-based discretization. Shape-function interpolation, algebraic derivatives, and element-level Gaussian quadrature address the first three challenges. 
The fourth challenge of spectral bias is addressed by investigating the efficacy of alternative network architectures within this framework in a comparative study of \glspl{mlp} with \gls{rff} mapping, \glspl{kan}, and \glspl{rbfnn}.

The remainder of this paper is organized as follows: Section~\ref{sec:nem} outlines the continuum mechanics foundation, the general \gls{nem} formulation, and the examined neural network architectures. Section~\ref{sec:mnem} presents the methodology of the \gls{mnem} and details how mesh-based discretization benefits the training process. Section~\ref{sec:case_studies} validates the approach through benchmark problems involving stiff inclusions and real-world material composites. Finally, Section~\ref{sec:conclusion} summarizes the findings and discusses future directions.

\section{Neural energy method}
\label{sec:nem}
The \gls{nem}, a generalization of the \gls{dem}~\cite{samaniego2020,nguyenthanh2020}, approaches the solution of boundary value problems as a minimization task governed by variational principles. Instead of assembling and solving a system of algebraic equations, the method utilizes a neural network to parameterize the continuous displacement field. The physical equilibrium state is subsequently determined by optimizing the network weights to minimize the system's total potential energy using gradient-based optimization.

\subsection{Continuum mechanics formulation}
To solve a \gls{bvp} in solid mechanics, let a body $\mathcal{B}$ be bounded by $\partial \mathcal{B}$ in the reference configuration. The motion of the body is described by the deformation mapping $\B \varphi(\B X)$ that maps material points from the reference configuration $\B X$ to the current configuration $\B x = \B \varphi(\B X)$~\cite{holzapfel2000}. The displacement field is defined as
\begin{align}
    \B u(\B X) = \B \varphi(\B X) - \B X.
\end{align} 
The primary kinematic measure is the deformation gradient $\B F$, given by
\begin{align}
    \B F = \frac{\partial \B \varphi(\B X)}{\partial \B X} = \B I + \frac{\partial \B u(\B X)}{\partial \B X}.
    \label{eq:deformation_gradient}
\end{align}
For a hyperelastic material governed by a strain energy density function $\psi(\B F)$, the first Piola-Kirchhoff stress tensor $\B P$ is defined as the conjugate stress to $\B F$,
\begin{align}
    \B P = \frac{\partial\psi}{\partial\B F}.
    \label{eq:piola_kirchhoff}
\end{align}
The static equilibrium condition requires that
\begin{align}
    \mathrm{div}\, \B P + \B b = \B 0 \quad \text{in } \mathcal{B},
    \label{eq:force_equilibrium}
\end{align}
subject to the boundary conditions
\begin{align}
    \B u &= \overline{\B u} \quad \text{on } \partial \mathcal{B}_u, \\
    \B P \cdot \B N &= \overline{\B t} \quad \text{on } \partial \mathcal{B}_t,
    \label{eq:bcs}
\end{align}
where $\B b$ denotes body forces, $\B N$ is the outward unit normal in the reference configuration, and $\overline{\B u}$ and $\overline{\B t}$ are prescribed displacements and tractions on the Dirichlet ($\partial \mathcal{B}_u$) and Neumann ($\partial \mathcal{B}_t$) boundaries, respectively. The Cauchy stress tensor can be computed from the first Piola-Kirchhoff stress as
\begin{align}
    \B \sigma = J^{-1} \B P \B F^T.
\end{align}

Rather than solving the strong form directly, the \gls{nem} seeks the state that minimizes the total potential energy functional $\Pi$,
\begin{align}
    \Pi(\B u) = \int_{\mathcal{B}} \psi(\B F)\:dV - \int_{\mathcal{B}} \B b \cdot \B u\:dV - \int_{\partial \mathcal{B}_t} \overline{\B t} \cdot \B u \: dA.
    \label{eq:potential_energy_whole}
\end{align}

For a heterogeneous composite of $n$ distinct material phases occupying subdomains $\mathcal{B}_1$, ..., $\mathcal{B}_n$, such that
\begin{align}
    \mathcal{B} = \bigcup_{i=1}^{n} \mathcal{B}_i \quad \text{with} \quad \mathcal{B}_i \cap \mathcal{B}_j = \emptyset \quad \forall i \neq j,
\end{align}
the energy functional can be written as
\begin{align}
    \Pi(\B u) = \sum_{i=1}^n \int_{\mathcal{B}_i} \psi_i(\B F)\:dV - \int_{\mathcal{B}} \B b \cdot \B u\:dV - \int_{\partial \mathcal{B}_t} \overline{\B t} \cdot \B u \: dA,
    \label{eq:potential_energy}
\end{align}
with the strain energy given per-material as $\psi_1$, ..., $\psi_n$.

\subsection{Network implementation and optimization}
In the \gls{nem}, the displacement field is parameterized by a neural network $\hat{\B u}(\B X; \B \theta)$ with parameters $\B \theta$. During training, the network takes a set of discrete points $\B X$ within $\mathcal{B}$ as input and predicts their displacements $\hat{\B u}$. The optimal parameter values $\B \theta^*$ are identified by minimizing the discrete potential energy,
\begin{align}
    \B \theta^* = \underset{\B \theta}{\textrm{arg min}}\:\Pi(\hat{\B u}(\B X; \B \theta)).
\end{align}

To strictly satisfy Dirichlet boundary conditions, the raw network output $\hat{\B z}(\B X; \B \theta)$ is transformed using a distance-function approach,
\begin{align}
    \hat{\B u}(\B X; \B \theta) &= \B z_P(\B X) + \B z_D(\B X) \odot \hat{\B z}(\B X; \B \theta),
\end{align}
where $\B z_P$ is a particular solution satisfying $\overline{\B u}$ on $\partial \mathcal{B}_u$, $\B z_D$ is a distance function that vanishes on $\partial \mathcal{B}_u$, and $\odot$ denotes the Hadamard product~\cite{sukumar2022}. 

\begin{figure}[!ht]
    \centering
    \input{tikz_nem_flowchart}
    \caption{Information flow in the \gls{nem}. Gray boxes indicate the components replaced in the \gls{mnem}.}
    \label{fig:dem_process}
\end{figure}
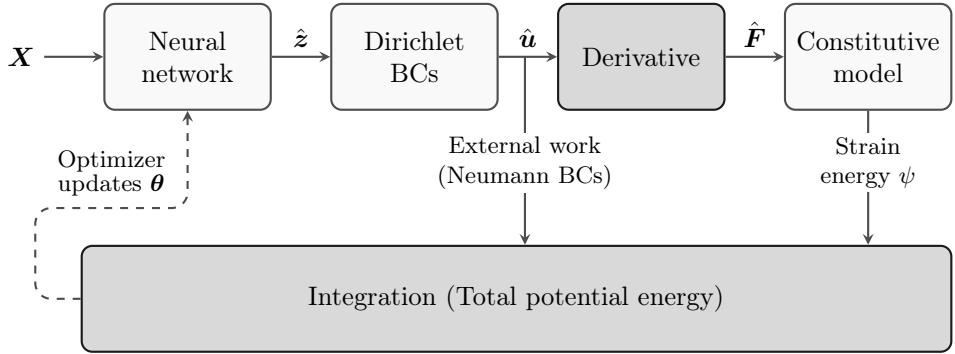

The deformation gradient is then computed by differentiating $\hat{\B u}$ with respect to the spatial coordinates. The strain energy density is evaluated via the respective constitutive law and integrated over the domain $\mathcal{B}$, while external forces are integrated over the boundary $\partial\mathcal{B}$, to obtain the total potential energy $\Pi$. Finally, the gradient $\partial \Pi / \partial \B \theta$ is computed via automatic differentiation and passed to the optimizer. The optimization is typically carried out using first-order gradient-based algorithms such as the Adam optimizer~\cite{kingma2017}, second-order quasi-Newton methods such as the \gls{lbfgs} algorithm~\cite{liu1989}, or a combination of both. The overall information flow is depicted in Fig.~\ref{fig:dem_process}. 

In the conventional implementation, referred to in the current work as the \gls{cnem}, the energy functional is evaluated on a set of collocation points and the deformation gradient is computed via automatic differentiation. This introduces three limitations that become critical for the simulation of heterogeneous composites. A sufficiently expressive network may produce physically inadmissible displacement oscillations between collocation points that remain invisible to the loss function as they do not affect the energy evaluated at the sampling locations, a phenomenon here referred to as geometric overfitting. Furthermore, computing the deformation gradient via automatic differentiation grows the computational graph with network depth and the number of integration points, increasing memory consumption and sensitivity to floating point precision. Finally, stochastic or low-order integration schemes introduce aliasing errors when the strain energy density varies abruptly across material interfaces.

\subsection{Alternative neural network architectures}
\label{sec:architectures}
The use of standard \glspl{mlp} in the \gls{nem} faces challenges in approximating functions with large spatial gradients, such as those occurring at phase boundaries in heterogeneous structures. This is due to spectral bias, a phenomenon where low-frequency features are learned effectively while high-frequency features are neglected~\cite{tancik2020}. Therefore, the current work investigates three alternative architectures which address spectral bias through different mechanisms: an \gls{mlp} with \gls{rff} mapping, a \gls{kan}, and an \gls{rbfnn}.

\subsubsection{Random Fourier feature mapping}
\gls{rff} mapping mitigates spectral bias by projecting the input coordinates into a higher-dimensional feature space before passing them to the network. This technique has shown enhanced performance in \glspl{pinn}~\cite{wang2021} and the \gls{cnem}~\cite{chadha2023}. The input $\B X \in \mathbb{R}^N$ is transformed via the mapping $\gamma: \mathbb{R}^N \to \mathbb{R}^{2M}$,
\begin{align}
    \gamma(\B X) = 
    \begin{bmatrix}
        \cos(2\pi\B B\B X) \\
        \sin(2\pi\B B\B X)
    \end{bmatrix},
\end{align}
where $\B B \in \mathbb{R}^{M \times N}$ is a matrix of weights sampled from a Gaussian distribution $\mathcal{N}(0, \sigma_{RFF}^2)$. The scale parameter $\sigma_{RFF}$ is a critical hyperparameter that controls the frequency spectrum of the input encoding. For brevity, the current work refers to an \gls{mlp} with \gls{rff} mapping applied to the input features as \gls{mlp}+\gls{rff}.

\subsubsection{Kolmogorov-Arnold networks}
\glspl{kan}~\cite{liu2025} are based on the Kolmogorov-Arnold representation theorem. While \glspl{mlp} rely on linear weight matrices coupled with fixed activation functions, \glspl{kan} employ learnable non-linear activation functions along the edges. A \gls{kan} layer maps an input vector $\B z^{(l)}$ to an output $\B z^{(l+1)}$ via a sum of univariate functions,
\begin{align}
    z_{j}^{(l+1)} = \sum_{i} \phi_{i,j}^{(l)}(z_i^{(l)}),
\end{align}
where $\B z^{(0)} = \B X$ corresponds to the input coordinates, and $\phi_{i,j}$ are typically parameterized as B-splines. This architecture allows the network to learn both the connections and the non-linearities simultaneously. Although \glspl{kan} are computationally more expensive per parameter than \glspl{mlp}, recent applications in physics-informed machine learning suggest that they offer superior accuracy and parameter efficiency for solving differential equations~\cite{wang2025kan}.

\subsubsection{Radial-basis-function neural networks}
An \gls{rbfnn} approximates the solution as a linear combination of radial basis functions, as suggested for the \gls{sem}~\cite{liang2022}. Unlike \glspl{mlp} which utilize global activation functions, \glspl{rbf} exhibit a localized response, which can lead to better conditioning and faster convergence for problems with localized deformation features.

The network structure consists of an input layer, a single hidden layer of RBF units, and a linear output layer. The approximation takes the form
\begin{align}
    \hat{\B z}(\B X) = \sum_{i=1}^{N_c} \B w_i \cdot \Phi\left( ||\B X - \B c_i|| \right) + \B b,
\end{align}
where $N_c$ is the number of kernel centers with coordinates $\B c_i$, $\B w_i$ are learnable weight vectors, $\B b$ is a learnable bias vector, and $\Phi$ is the radial basis function.
While the kernel centers can be unstructured, a uniform distribution on a rectangular domain allows the definition of a linear center density $k$ as
\begin{align}
    k = \frac{k_x-1}{L_x} = \frac{k_y-1}{L_y},
    \label{eq:kerneldensity}
\end{align}
where $L_x$, $L_y$ are the dimensions of the domain and $k_x$, $k_y$ denote the number of \gls{rbf} kernel centers along the respective axes.

The most effective kernel for the \gls{sem} is the Matérn-$\frac{3}{2}$ function~\cite{liang2022}. It is $C^1$ continuous, satisfying the continuity requirements of the weak form, and is defined as
\begin{align}
    \Phi^{M\frac{3}{2}}(r_i) &= \left(1+\frac{\sqrt{3}r_i}{v_i}\right)
    \exp\left(-\frac{\sqrt{3}r_i}{v_i}\right),
    \label{eq:matern32}
\end{align}
where $r_i = ||\B X - \B c_i||$ is the Euclidean distance, and $v_i$ is the width of the $i^\textrm{th}$ center. To ensure sufficient kernel overlap, $v_i$ is typically initialized based on the local grid density,
\begin{align}
    v_i = \frac{d_i^{max}}{\sqrt{2\min(k_x,~k_y)}},
\end{align}
with $d_i^{max}$ denoting the maximum distance between the $i^\textrm{th}$ kernel center and the other kernel centers. 

As an enhancement to the \gls{sem}, the current work proposes a modification to the widths of the \glspl{rbf}. Each shape parameter $v_i$ is scaled by a scalar length scale $l$,
\begin{align}
    \tilde{v}_i = l \cdot v_i,
\end{align}
which acts as an additional hyperparameter. It can be optimized in a pre-processing step by minimizing the condition number of the \gls{rbf} kernel matrix. This strategy allows for the efficient determination of the optimal shape parameter without the computational expense of training the network for every trial value in a hyperparameter optimization study.

Note that in the current work, the kernel center positions are fixed and not trainable. Therefore, when training the \gls{rbfnn}, only the weights and bias of the output layer are optimized.

\section{Mesh-based neural energy method}
\label{sec:mnem}
The current work presents the \gls{mnem}, extending the \gls{nem} by introducing shape functions for the interpolation of the displacement field and the computation of strains, and Gaussian quadrature the numerical integration of the energy functional. This addresses the limitations of the \gls{cnem} identified in Section~\ref{sec:nem}. 
Shape function interpolation poses a kinematic constraint, preventing the sub-sampling oscillations that arise in the \gls{cnem}. Moreover, it allows the algebraic computation of derivatives, avoiding the cost of automatic differentiation and stabilizing training on heterogeneous materials. Finally, integration by Gaussian quadrature eliminates aliasing errors.

\subsection{Domain discretization and interpolation}
In the \gls{mnem}, the body $\mathcal{B}$ is discretized into $n_{el}$
isoparametric elements~\cite{wriggers2008} defined by a set of global nodes $\mathcal{N} = \{ \B X_1, \B X_2, \dots, \B X_{n_{node}} \}$. The neural network maps the spatial coordinates of these nodes to their corresponding displacement vectors
\begin{align}
    \hat{\B u}_a = \hat{\B u}(\B X_a; \B \theta),
\end{align}
for all nodes $a \in [1, ..., n_{node}]$, where $\B \theta$ represents the trainable parameters of the network. 

Following the standard isoparametric concept, the geometry $\B X$ and the displacement field $\B u$ within an element domain $\mathcal{B}^e$ are interpolated using the same set of shape functions $N_a(\B \xi)$ defined in the parent coordinates $\B \xi$,
\begin{align}
    \B X(\B \xi) &= \sum_{a=1}^{n_{en}} N_a(\B \xi) \B X_a^e, \\
    \B u^e(\B \xi) &= \sum_{a=1}^{n_{en}} N_a(\B \xi) \hat{\B u}_a^e,
    \label{eq:displacement_interpolation}
\end{align}
where $n_{en}$ is the number of nodes per element, and $\B X_a^e$ and $\hat{\B u}_a^e$ are the coordinates and network-predicted displacements of the element nodes, respectively. 

The spatial derivatives of the displacement field are computed algebraically as
\begin{align}
    \frac{\partial \B u^e}{\partial \B X} = \sum_{a=1}^{n_{en}} \hat{\B u}_a^e \otimes \nabla_{\B X} N_a(\B \xi),
\end{align}
where $\nabla_{\B X} N_a = \B J^{-T} \nabla_{\B \xi} N_a$, and $\B J$ is the geometric Jacobian of the mapping. The deformation gradient is then assembled as
\begin{align}
    \B F(\B \xi) = \B I + \frac{\partial \B u^e}{\partial \B X}.
\end{align}
This approach decouples the computation of spatial derivatives from the neural network's automatic differentiation graph, which is now reserved solely for optimizing the parameters $\B \theta$. 

\subsection{Numerical integration and optimization}
\label{sec:integration}

The required numerical integrations are performed using Gaussian quadrature~\cite{wriggers2008}. The integral of a generic function $f(\B X)$ over the element is approximated by
\begin{align}
    \int_{\mathcal{B}^e} f(\B X) \, dV \approx \sum_{k=1}^{n_{gp}} w_k f(\B \xi_k) \det\B J(\B \xi_k),
\end{align}
where $n_{gp}$ is the number of integration points, $w_k$ are the quadrature weights, and $\B \xi_k$ are the integration point coordinates in the parent domain. The term $\det\B J$ represents the determinant of the Jacobian of the isoparametric mapping at the integration point. 

Consequently, the total potential energy $\Pi$ (Eq.~\eqref{eq:potential_energy}) is approximated by 
\begin{align}
    \Pi(\B \theta) \approx \sum_{e=1}^{n_{el}} \sum_{k=1}^{n_{gp}} w_k \det\B J(\B \xi_k) \left[ \psi \left( \B F(\B \xi_k) \right) - \B b(\B X_k) \cdot \B u^e(\B \xi_k) \right] - W_{ext}.
    \label{eq:qdem_energy}
\end{align}
Here, $W_{ext}$ denotes the external work done by surface tractions, computed via the trapezoidal rule as
\begin{align}
    W_{ext} = \int_{\partial \mathcal{B}_t} \overline{\B t} \cdot \B u \: dA \approx \sum_{j=1}^{n_{bt}} \frac{l_j}{2} \left( \overline{\B t}_j \cdot \B u_j + \overline{\B t}_{j+1} \cdot \B u_{j+1} \right),
    \label{eq:wext}
\end{align}
where $n_{bt}$ is the number of boundary segments on $\partial \mathcal{B}_t$, $l_j$ is the length of the $j^\textrm{th}$ segment, and subscripts $j$, $j+1$ denote the segment endpoints.

In summary, the information flow in the \gls{mnem} proceeds as follows: First, the global node coordinates are passed through the neural network to obtain the displacements $\hat{\B u}_a$. Then, these nodal values are broadcast to the integration points via Eq.~\eqref{eq:displacement_interpolation} to compute spatial gradients using shape functions. The strain energy density $\psi$ is evaluated at the integration points and integrated over the domain $\mathcal{B}$, while the external work is evaluated at boundary nodes and integrated over the boundary $\partial\mathcal{B}$, to compute the total potential energy $\Pi$. Finally, the gradient of the total potential energy with respect to the network parameters, $\partial \Pi / \partial \B \theta$, is computed via automatic differentiation and passed to the optimizer to update the network weights. The components exchanged relative to the \gls{cnem} are highlighted in gray in Fig.~\ref{fig:dem_process}.

Notably, the \gls{mnem} formulation is particularly advantageous for the \gls{rbfnn} architecture. In the \gls{cnem}, computing the deformation gradient requires differentiating the displacement field with respect to the spatial coordinates via automatic differentiation, which involves evaluating derivatives of the radial basis functions. The resulting computational graph is expensive to construct and memory-intensive to store. In the \gls{mnem}, the deformation gradient is instead obtained from algebraic shape function derivatives of the interpolated nodal displacements, eliminating this cost entirely. The automatic differentiation graph is then limited to the gradient of the loss with respect to the trainable output weights, which is trivial for the linear output layer. This not only reduces runtime but also alleviates a memory bottleneck, allowing the \gls{mnem} to scale to much higher kernel densities than the \gls{cnem} before encountering hardware limits.

\section{Numerical experiments}
\label{sec:case_studies}

The performance of the \gls{cnem} and \gls{mnem} is assessed through comparative case studies on (i) a rectangular coupon with a hard inclusion, (ii) a unit cell of the microstructure of a calcified hydrogel, and (iii) a volume element of a \gls{cfrp} microstructure.

\subsection{Experimental setup}

The \glspl{nem} are implemented in PyTorch 2.7.1~\cite{paszke2019}. All computations are performed in single-precision floating-point arithmetic (FP32) on an NVIDIA GeForce RTX 4070 Ti \gls{gpu} with 12 GB of VRAM. Preliminary experiments showed that switching to double precision (FP64) increased computation time and memory consumption noticeably without yielding meaningful gains in accuracy.

For each configuration, 30 training runs are completed with different random seeds. Because the loss is a global integral over the entire physical domain, training is conducted using full-batch gradient descent rather than stochastic mini-batching. Optimization is performed in two stages: a pre-training phase with the Adam optimizer for 500 epochs at a step size of $1\times10^{-4}$ provides initial stability, followed by fine-tuning with the \gls{lbfgs} algorithm~\cite{liu1989}.

For hyperelastic constitutive models with a logarithmic volumetric penalty term $\ln(J)$, the admissibility condition $J>0$ must be maintained throughout training. If an intermediate prediction yields a nonphysical deformation with $J\leq0$, the natural logarithm is undefined and the computational graph collapses. To prevent this, $J$ is clamped to a small positive lower bound of $10^{-6}$ during the forward pass.

To ensure a consistent evaluation of computational effort and solution quality, an automated convergence criterion is employed. Adapting the approach from \cite{he2023elastoplasticity}, convergence is defined as
\begin{align}
    \left| \frac{\frac{1}{N}\sum_{i=0}^{N-1}\mathcal{L}_{j-i} - 
    \frac{1}{N}\sum_{i=N}^{2N-1}\mathcal{L}_{j-i}}
    {\frac{1}{N}\sum_{i=0}^{N-1}\mathcal{L}_{j-i}} \right| \leq \tau_{tol},
    \label{eq:conv}
\end{align}
where $\mathcal{L}_j$ denotes the loss value at the $j^{\text{th}}$ epoch, $N$ represents the window size for the moving average, and $\tau_{tol}$ is the user-defined tolerance. For this work, a window size of $N=20$ and a strict tolerance of $\tau_{tol}=1\times10^{-8}$ are used.

Hyperparameter choices are kept consistent across architectures where possible. For the \gls{mlp} and \gls{kan} architectures, Tanh is used as the activation function throughout, since it is well established in the physics-informed literature~\cite{raissi2019, samaniego2020} for providing the smooth, continuous, and nonzero spatial derivatives required to construct valid strain fields. Where \gls{rff} mapping is applied, the standard deviation $\sigma_{RFF}$ governing the frequency spectrum of the input encoding is set to 1. The \gls{kan} parameters follow the recommendations of the foundational \gls{kan} literature~\cite{liu2025, wang2025kan}, and the \gls{rbfnn} uses the Matérn-$\frac{3}{2}$ kernel, following the recommendation for the \gls{sem}~\cite{liang2022}. For each case study, the \gls{rbfnn} kernel length scale is optimized by a grid search that minimizes the condition number of the kernel matrix. The network sizes are scaled according to the problem complexity.

To validate the accuracy of the proposed methods, results are compared against \gls{fem} reference solutions computed with FEniCSx (DOLFINx 0.9.0~\cite{dolfinx}, Basix 0.9.0~\cite{basix1,basix2}, UFL 2024.2.0~\cite{ufl}). As a scalar measure for global accuracy, the relative $L^2$ stress error with respect to the reference solution, evaluated over the entire physical domain $\mathcal{B}$ from the Cauchy stress tensor $\hat{\B \sigma}$, is calculated as
\begin{align}
    \epsilon_{\B \sigma} = \frac{||\hat{\B \sigma} - \B \sigma_{FEM}||_{L^2}}{||\B \sigma_{FEM}||_{L^2}},
\end{align}
where the $L^2$ norm of a tensor field is computed via
\begin{align}
    ||\B \sigma||_{L^2} = \sqrt{\int_{\mathcal{B}}||\B \sigma||_2^2\:dV}.
\end{align}
The integrals are evaluated using the same quadrature rule applied for the integration of internal and external energy. Stresses are utilized for the $L^2$ error, as they provide a more sensitive measure of mechanical accuracy than displacements. Stress plots show the solution of the run with the median stress error.

\subsection{Constitutive models}
\label{sec:constitutive_models}
To characterize the mechanical response of the materials investigated in this work, different constitutive formulations are required. Two distinct classes of material behavior are employed: linear elasticity for small-deformation problems, and hyperelasticity for applications involving large deformations and nonlinear material response.

\subsubsection{Linear elasticity}
For small deformations of isotropic linear elastic solids, the strain energy density according to Hooke's law is given by
\begin{align}
    \psi(\B \varepsilon) = \frac{1}{2} \lambda (\mathrm{tr} \left\{\B \varepsilon\right\})^2 + \mu \,\, \B \varepsilon : \B \varepsilon,
\end{align}
where $\B \varepsilon = \frac{1}{2}(\nabla \B u + \nabla \B u^T)$ is the linearized strain tensor, and $\lambda = \frac{E \nu}{(1+\nu)(1-2\nu)}$ and $\mu = \frac{E}{2(1+\nu)}$ are the Lamé parameters with Young's modulus $E$ and Poisson's ratio $\nu$. The Cauchy stress tensor is computed as
\begin{align}
    \B \sigma = 2 \mu \B \varepsilon + \lambda \,\, \mathrm{tr} \left\{\B \varepsilon\right\} \B I.
\end{align}

\subsubsection{Hyperelasticity}
In isotropic hyperelasticity, the three invariants of the right Cauchy-Green deformation tensor $\B C = \B F^T \B F$ are
\begin{align}
    I_1 &= \lambda_1^2 + \lambda_2^2 + \lambda_3^2, \\
    I_2 &= \lambda_1^2\lambda_2^2 + \lambda_2^2\lambda_3^2 + \lambda_3^2\lambda_1^2, \\
    J &= \lambda_1 \lambda_2 \lambda_3
\end{align}
with the principal stretches $\lambda_1, \lambda_2, \lambda_3$. For two-dimensional problems, only $\lambda_1$ and $\lambda_2$ are obtained from $\B C$. In a plane strain case the third principal stretch is set to $\lambda_3=1$, and in an incompressible plane stress case it is computed by $\lambda_3 = (\lambda_1 \lambda_2)^{-1}$.

\paragraph{Neo-Hookean model}
A compressible formulation with a logarithmic volumetric penalty is used, given by
\begin{align}
    \psi_{NH} = \frac{\mu}{2} (I_1 - 3) - \mu \ln J + \frac{\lambda}{2} (\ln J)^2.
\end{align}

\paragraph{Mooney-Rivlin model}
The energy density of the Mooney-Rivlin model takes the form
\begin{align}
    \psi_{MR} = C_{10}(I_1 - 3) + C_{01}(I_2 - 3),
\end{align}
where $C_{10}$ and $C_{01}$ are material constants related to the initial shear modulus by $\mu = 2(C_{10} + C_{01})$.

\paragraph{Ogden model}
The energy density of the incompressible Ogden model is given by
\begin{align}
    \psi_{Ogden} = \sum_{i=1}^M \frac{\mu_i}{\alpha_i} \left( \lambda_1^{\alpha_i} + \lambda_2^{\alpha_i} + \lambda_3^{\alpha_i} - 3 \right),
\end{align}
where $M$ is the model order and $\mu_i$ and $\alpha_i$ are material parameters.

\subsection{Coupon with stiff inclusion}
\label{sec:coupon}
\subsubsection{Problem definition}
\begin{figure}[!ht]
    \centering
    \input{tikz_coupon}
    \caption{Geometry of the coupon with stiff inclusion.}
    \label{fig:coupon}
\end{figure}
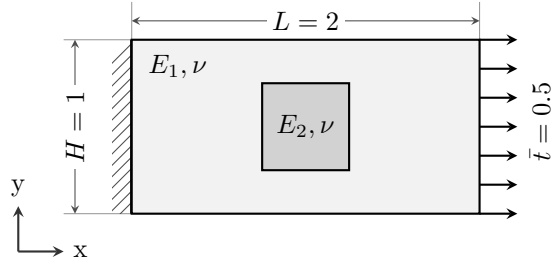

The first case study examines a heterogeneous coupon consisting of two materials with varying stiffness, as illustrated in Fig.~\ref{fig:coupon}. The rectangular domain has length $L=2$ and height $H=1$. The boundary conditions mimic a standard tensile test: the left boundary is fully clamped, the right boundary is subjected to a uniform surface traction of $\overline{t}=0.5$ in the positive $x$-direction, and the upper and lower boundaries are traction-free. Both materials are modeled as a neo-Hookean solid under plane strain conditions, where the inclusion has a stiffness ten times higher than the surrounding matrix ($E_1=1$, $E_2=10$, $\nu=0.3$). This example serves as a numerical benchmark to assess the quality of the \gls{mnem} compared to the \gls{cnem}, where all quantities are dimensionless.

For the \gls{cnem}, the domain is discretized by 3,321 collocation points and integration is performed by the trapezoidal rule. The coordinates of these points are used as nodes for the \gls{mnem} and \gls{fem} to construct a structured mesh of 3,200 first-order quadrilateral elements with full Gauss integration.

Both \gls{cnem} and \gls{mnem} are evaluated using \gls{mlp}+\gls{rff}, \gls{kan}, and \gls{rbfnn} network architectures. The hyperparameters for each architecture are listed in Table~\ref{tab:hp_coupon}. Although recent literature recommends grid sizes up to 100 for \gls{kan}s in heterogeneous problems~\cite{wang2025kan}, preliminary experiments with the standard \gls{cnem} show that such high resolutions induce a severe optimization instability. Consequently, the \gls{kan} grid size is restricted to 20. Moreover, the kernel resolution of the \gls{rbfnn} is limited to 50 to prevent excessive memory consumption exceeding the available hardware limits, caused by the backpropagation of the automatic differentiation graph through the \gls{rbf} layer in the \gls{cnem}. While the \gls{mnem} is not subject to these stability and memory bottlenecks, identical hyperparameters are chosen for both methods to maintain straightforward comparability. The modified kernel length scale of $l=0.06$ induces divergence in the \gls{cnem}, which therefore requires a fallback to $l=1$ to remain stable; the \gls{mnem} is trained with the modified value throughout. Despite identical network sizes, the training dynamics with the \gls{lbfgs} algorithm differ significantly: the \gls{cnem} requires a reduced step size of $1\times10^{-4}$ to prevent divergence, whereas the \gls{mnem} maintains stability with a step size of 1, resulting in faster convergence.

\begin{table}[ht]
    \centering
    \caption{Hyperparameter values used for the coupon with stiff inclusion.}
    \label{tab:hp_coupon}
    \begin{tabular}{lccc}
         \toprule
         \textbf{Hyperparameter} & \textbf{\gls{mlp}+\gls{rff}} & \textbf{\gls{kan}} & \textbf{\gls{rbfnn}} \\ 
         \midrule
         \textit{General parameters} & & & \\
         Number of hidden layers     & 10   & 2    & 1 \\ 
         Dimension of hidden layers  & 64   & 32   & -- \\ 
         Activation / Kernel function & Tanh & Tanh & Matérn-$\frac{3}{2}$ \\ 
         \midrule
         \textit{Discretization \& Encoding} & & & \\
         \gls{rff} mapping std. dev.     & 1    & --   & -- \\ 
         Spline order                & --   & 3    & -- \\ 
         Grid / Kernel Resolution    & --   & 20   & 50 \\ 
         Kernel length scale         & --   & --   & 0.06 \\ 
         \bottomrule
    \end{tabular}
\end{table}

\subsubsection{Results}
\begin{figure}[!ht]
    \centering
    \includegraphics{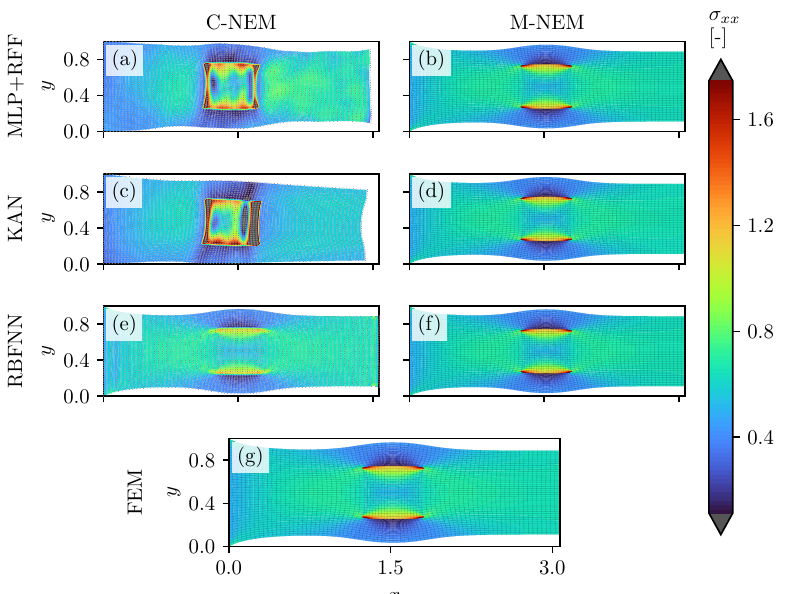}
    \caption{Normal stress ($\sigma_{xx}$) solutions for the coupon with stiff inclusion: (a,c,e) the \gls{cnem} and (b,d,f) the \gls{mnem} with different neural network architectures, compared to (g) the \gls{fem} reference solution.}
    \label{fig:coupon_sigma_xx}
\end{figure}

Fig.~\ref{fig:coupon_sigma_xx} illustrates the normal stress distribution in the $x$-direction for all examined combinations. Colorbars are scaled to the range of values from the FEM reference solution, with values outside this range being rendered in gray to highlight deviations. While the displacement field is approximated well by both methods, the stress predictions differ substantially. The \gls{cnem} exhibits significant oscillatory artifacts and overshoots stress values at the material interfaces. In contrast, the \gls{mnem} correctly captures the smooth stress distribution observed in the reference solution.

\begin{figure}[!ht]
    \centering
    \includegraphics{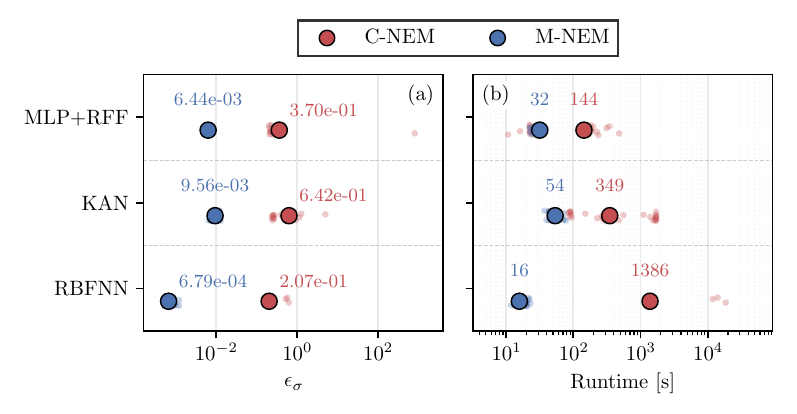}
    \caption{Comparison of log-scaled relative  stress error (a) and runtime (b) for the \gls{cnem} and \gls{mnem} with different neural network architectures on the coupon with stiff inclusion. Large dots show the median of the 30 training runs.}
    \label{fig:coupon_comparison}
\end{figure}

The quantitative performance comparison is presented in Fig.~\ref{fig:coupon_comparison}. The \gls{mnem} yields stress errors around two orders of magnitude lower than the \gls{cnem} while being faster across all architectures. This efficiency gain is attributed to both improved training stability, which reduces the required number of epochs, and a decrease in the computational cost per epoch. Here, the slightly higher cost of the Gaussian quadrature implementation is offset by the more efficient computation of derivatives using shape functions instead of automatic differentiation. Beyond the reduction in stress error and runtime, the \gls{mnem} also produces a tighter spread of both quantities across the 30 training runs, indicating a more reliable optimization process.

It is noteworthy that distinct performance differences arise among the network architectures in the \gls{mnem}, which are less pronounced within the \gls{cnem} framework. Contrary to the theoretical expressivity advantage of B-splines, the \gls{mlp}+\gls{rff} achieves slightly lower errors than the \gls{kan} in this benchmark. The restricted grid resolution imposed on the \gls{kan} to ensure optimization stability may contribute to this result, as the ranking is not observed consistently across the case studies, as shown in Section~\ref{sec:hydrogel}. The \gls{rbfnn} demonstrates superior performance across all metrics. It achieves the lowest approximation error while simultaneously requiring the least computational time, outperforming the \gls{mlp}+\gls{rff} in both accuracy and speed.
As expected, the \gls{rbfnn} architecture benefits most significantly from the transition to the \gls{mnem} framework, exhibiting the largest relative reduction in runtime. 

\begin{figure}[!ht]
    \centering
    \includegraphics{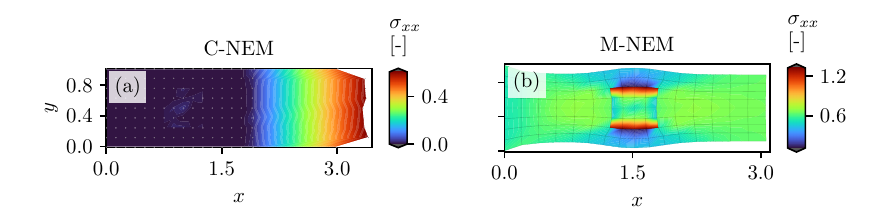}
    \caption{Normal stress ($\sigma_{xx}$) solutions for \gls{cnem} (a) and \gls{mnem} (b) on the coupon with stiff inclusion with a coarse discretization and \gls{mlp}+\gls{rff} architecture.}
    \label{fig:coupon_coarse}
\end{figure}

Finally, beyond computational speed, the \gls{mnem} demonstrates superior robustness regarding hyperparameter selection and domain discretization. While the \gls{cnem} often necessitates careful manual tuning to prevent divergence on heterogeneous structures, the \gls{mnem} remains stable across a wider range of settings. Furthermore, the \gls{mnem} was observed to converge with coarser discretizations than those required by the \gls{cnem}. When the spatial resolution is reduced, the \gls{cnem} fails to converge: because the potential energy is evaluated only at the widely spaced collocation points, an expressive network can develop physically inadmissible deformations between them without this being penalized by the loss. As training proceeds, the optimizer drives the loss below the true potential energy, at which point the gradients explode and training diverges. As shown in Fig.~\ref{fig:coupon_coarse}, the \gls{cnem} training process on a coarse mesh is aborted at the onset of this instability, visible as a deformation field that tears the body apart between the collocation points, whereas the \gls{mnem} converges to a smooth, physically admissible solution under equivalent conditions. This robustness stems from the shape function interpolation, which kinematically constrains the solution space between nodes, combined with the Gaussian quadrature, which evaluates the energy over the full element volume rather than only at the nodes and thereby penalizes erratic deformations.

\subsection{Calcified hydrogel unit cell}
\label{sec:hydrogel}
\subsubsection{Problem definition}
The second case study applies the \gls{mnem} to a unit cell~\cite{zhao2025} of the microstructure of a calcified hydrogel. Calcified hydrogels are bioinspired soft–stiff composites designed to mimic the hierarchical structure of mineralized biological tissues, such as bone or nacre~\cite{rauner2017}. They consist of a hydrated polymer network reinforced by in situ mineral phases, combining large deformation capabilities with high stiffness. The resulting heterogeneous microstructure exhibits a highly nonlinear mechanical response with localized stress fields.

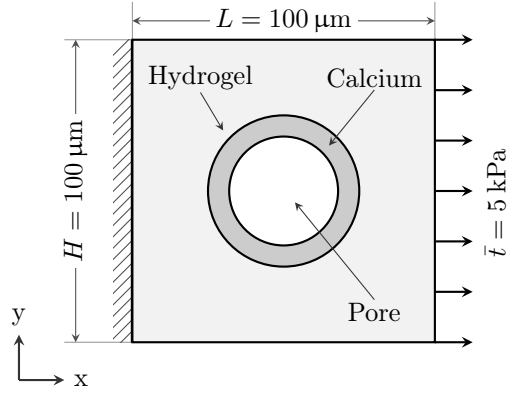
\begin{figure}[!ht]
    \centering
    \input{tikz_hydrogel}
    \caption{Idealized simulation domain of the calcified hydrogel unit cell problem.}
    \label{fig:unitcell_geometry}
\end{figure}

Based on microscopic observations~\cite{rauner2017,burchak2021} and numerical investigations~\cite{graham2024, ayguen2024}, this $100 \times 100\:\mathrm{\upmu m}$ unit cell consists of a \gls{pam} hydrogel matrix (80\% volume fraction) containing a circular, hollow calcium inclusion (10\% calcium, 10\% pore), as illustrated in Fig.~\ref{fig:unitcell_geometry}. The problem is modeled under incompressible plane stress conditions. The boundary conditions are analogous to the previous study, with a fully clamped left boundary and a uniform tensile traction of $\overline{t}=5\:\mathrm{kPa}$ on the right boundary.

\begin{figure}[!ht]
    \centering
    \includegraphics{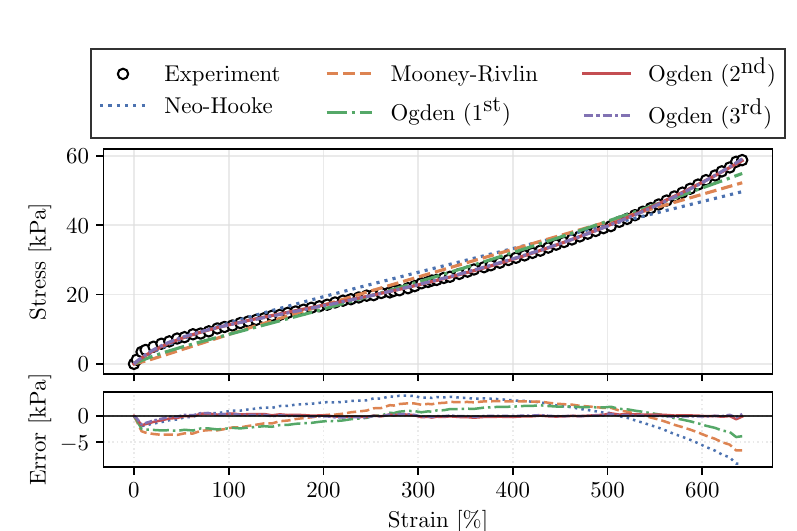}
    \caption{Fitting hyperelastic material models to experimental \gls{pam} hydrogel data from \cite{yang2025}.}
    \label{fig:yang_data}
\end{figure}

To identify the most suitable constitutive model for the hydrogel matrix, a neo-Hookean, a Mooney-Rivlin, and Ogden models of first through third order are calibrated via a nonlinear least squares fit~\cite{ogden2004} on experimental stress-strain data digitised from~\cite{yang2025}. As shown in Fig.~\ref{fig:yang_data}, the standard neo-Hookean and Mooney-Rivlin models fail to capture the stiffening behavior at larger strains. Consequently, the second-order Ogden model ($M=2$) is selected. The calibrated material parameters are summarized in Table~\ref{tab:unitcell_ogden_parameters}.
The calcium inclusion is modeled as a linear elastic solid with a Young's modulus of $2 \times 10^7\:\mathrm{Pa}$ and Poisson's ratio of $0.3$, resulting in a stiffness contrast of approximately three orders of magnitude.

\begin{table}[ht]
    \centering
    \caption{Material parameters for the \gls{pam} hydrogel matrix, derived via a two-term Ogden model fit.}
    \label{tab:unitcell_ogden_parameters}
    \begin{tabular}{ll c l}
         \toprule
         \textbf{Description} & \textbf{Symbol} & \textbf{Value} & \textbf{Unit} \\ 
         \midrule
         Shear modulus          & $\mu_1$    & $1 \times 10^4$ & Pa \\ 
         Nonlinearity parameter & $\alpha_1$ & $1.503$         & -- \\ 
         Shear modulus          & $\mu_2$    & $7.43$          & Pa \\ 
         Nonlinearity parameter & $\alpha_2$ & $4.07$          & -- \\ 
         \bottomrule
    \end{tabular}
\end{table}

The complexity of this setup arises from the combination of a high stiffness contrast and the topological intricacies introduced by the traction-free internal void. The domain is discretized by 8,313 first-order triangular elements with full integration for both \gls{mnem} and \gls{fem}. Due to the severe convergence difficulties encountered by the \gls{cnem} in the presence of such high stiffness ratios and nonlinear material behavior, only \gls{mnem} results are reported. The hyperparameters are adapted for sufficient expressivity and are given in Table~\ref{tab:hp_hydrogel}. 

\begin{table}[ht]
    \centering
    \caption{Hyperparameter values used for the calcified hydrogel unit cell.}
    \label{tab:hp_hydrogel}
    \begin{tabular}{lccc}
         \toprule
         \textbf{Hyperparameter} & \textbf{\gls{mlp}+\gls{rff}} & \textbf{\gls{kan}} & \textbf{\gls{rbfnn}} \\ 
         \midrule
         \textit{General parameters} & & & \\
         Number of hidden layers     & 10   & 2    & 1 \\ 
         Dimension of hidden layers  & 256   & 32   & -- \\ 
         Activation / Kernel function & Tanh & Tanh & Matérn-$\frac{3}{2}$ \\ 
         \midrule
         \textit{Discretization \& Encoding} & & & \\
         \gls{rff} mapping std. dev.     & 1    & --   & -- \\ 
         Spline order                & --   & 3    & -- \\ 
         Grid / Kernel Resolution    & --   & 200  & 200 \\ 
         Kernel length scale         & --   & --   & 0.12 \\ 
         \bottomrule
    \end{tabular}
\end{table}

\subsubsection{Results}
\begin{figure}[!ht]
    \centering
    \includegraphics{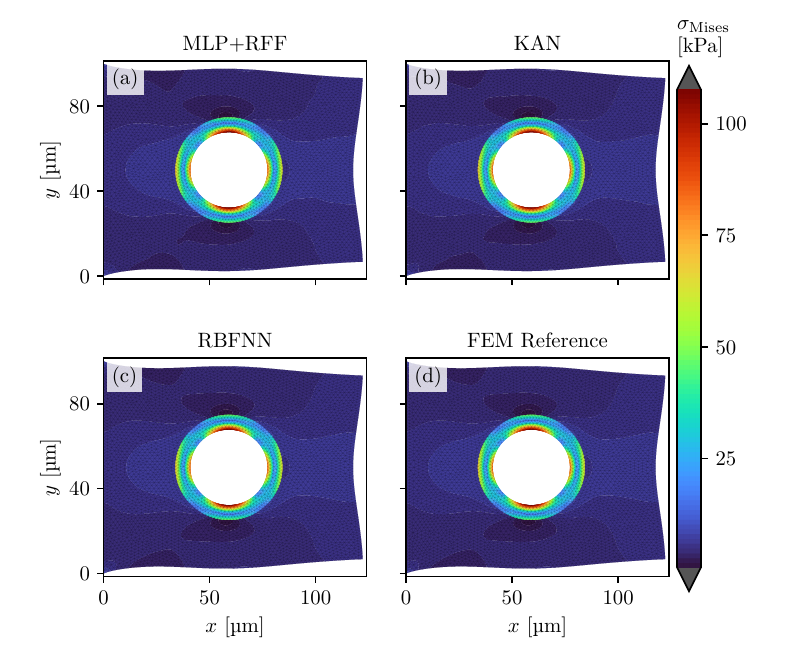}
    \caption{Von-Mises stress solutions for the calcified hydrogel unit cell: (a-c) the \gls{mnem} with different neural network architectures, and (d) the \gls{fem} reference solution.}
    \label{fig:unitcell_sigma}
\end{figure}

Fig.~\ref{fig:unitcell_sigma} shows the von-Mises stress distribution obtained with the \gls{mnem} for all three network architectures, alongside the \gls{fem} reference solution. All three variants yield a stress field that is visually indistinguishable from the reference solution, indicating that the \gls{mnem} accurately resolves both the displacement and stress response despite the severe stiffness contrast and the traction-free internal void.

\begin{figure}[!ht]
    \centering
    \includegraphics{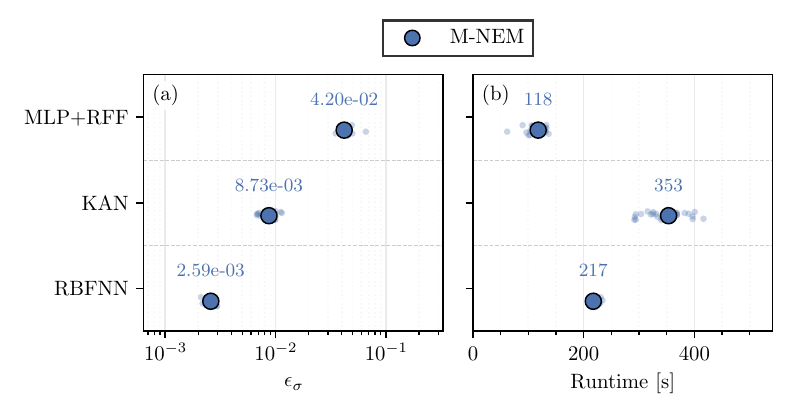}
    \caption{Comparison of log-scaled relative  stress error (a) and runtime (b) for the \gls{mnem} with different neural network architectures on the calcified hydrogel unit cell. Large dots show the median of the 30 training runs.}
    \label{fig:unitcell_comparison}
\end{figure}

The quantitative comparison of stress errors in Fig.~\ref{fig:unitcell_comparison} reveals differences in solution quality. In contrast to the coupon benchmark, where the \gls{mlp}+\gls{rff} matched or outperformed the \gls{kan}, the \gls{kan} here achieves a lower error, suggesting that the localized support of B-splines is better suited to capturing the complex geometry of the porous unit cell. The overall highest accuracy is again obtained by the \gls{rbfnn}, reaching an error approximately one order of magnitude lower than the \gls{mlp}+\gls{rff}. Regarding computational efficiency, the \gls{kan} exhibits the highest runtime of the three architectures, while the \gls{rbfnn} converges faster than the \gls{kan} but slower than the \gls{mlp}+\gls{rff}, offering the most favorable overall compromise between speed and accuracy.

\subsection{Carbon fiber reinforced polymer}
\label{sec:microstructure}
\subsubsection{Problem definition}
The third case study models a volume element of a unidirectional \gls{cfrp} spar cap, typically used in wind turbine blades. While the previous case study addressed extreme stiffness ratios and material nonlinearity, this example evaluates the method's performance on a domain with multiple inclusions. Furthermore, it tests the numerical stability of the \gls{mnem} for large absolute stiffness values, which typically presents a challenge for optimization without parameter scaling.

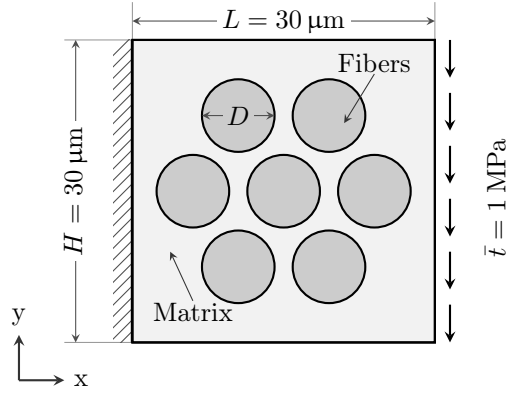
\begin{figure}[!ht]
    \centering
    \input{tikz_fiber}
    \caption{Idealized simulation domain of the \gls{cfrp}.}
    \label{fig:fiber_geometry}
\end{figure}

The geometric domain is a square $30\times30\:\mathrm{\upmu m}$ cell containing seven fibers with a diameter of $D=7.2\:\mathrm{\upmu m}$~\cite{zoltek2021}, arranged in a hexagonal packing pattern (see Fig.~\ref{fig:fiber_geometry}), modeled under plane strain conditions. The microstructure consists of Zoltek PX35 carbon fibers~\cite{zoltek2021} embedded in a \gls{pa66} matrix~\cite{kern}. The fibers are modeled as isotropic linear elastic solids with a Young's modulus of $E_f = 242\:\mathrm{GPa}$~\cite{zoltek2021} and a Poisson's ratio of $\nu_f = 0.2$~\cite{chawla2019}. 
The \gls{pa66} matrix is modeled in its conditioned state (equilibrium moisture content) to reflect realistic operating conditions. Since \gls{pa66} is inherently hygroscopic, moisture absorption significantly reduces stiffness while increasing toughness~\cite{shirinbayan2024}. Based on this, the material is modeled as a linear elastic solid with a Young's modulus of $E_m = 1.1\:\mathrm{GPa}$~\cite{kern} and a Poisson's ratio of $\nu_m = 0.4$~\cite{shirinbayan2024}.

The boundary conditions consist of a fully clamped left boundary and a uniform shear traction of $\overline{t}=1\:\mathrm{MPa}$ in the negative $y$-direction on the right boundary; the top and bottom boundaries are traction-free. The domain is discretized by 4,818 first-order triangular elements with full integration for both \gls{mnem} and \gls{fem}. Consistent with the previous section, results are reported exclusively for the \gls{mnem}, utilizing the hyperparameter configuration listed in Table~\ref{tab:hp_cfrp}.

\begin{table}[ht]
    \centering
    \caption{Hyperparameter values used for the \gls{cfrp}.}
    \label{tab:hp_cfrp}
    \begin{tabular}{lccc}
         \toprule
         \textbf{Hyperparameter} & \textbf{\gls{mlp}+\gls{rff}} & \textbf{\gls{kan}} & \textbf{\gls{rbfnn}} \\ 
         \midrule
         \textit{General parameters} & & & \\
         Number of hidden layers     & 10   & 2    & 1 \\ 
         Dimension of hidden layers  & 256   & 32   & -- \\ 
         Activation / Kernel function & Tanh & Tanh & Matérn-$\frac{3}{2}$ \\ 
         \midrule
         \textit{Discretization \& Encoding} & & & \\
         \gls{rff} mapping std. dev.     & 1    & --   & -- \\ 
         Spline order                & --   & 3    & -- \\ 
         Grid / Kernel Resolution    & --   & 100  & 200 \\ 
         Kernel length scale         & --   & --   & 0.06 \\ 
         \bottomrule
    \end{tabular}
\end{table}

\subsubsection{Results}
\begin{figure}[!ht]
    \centering
    \includegraphics{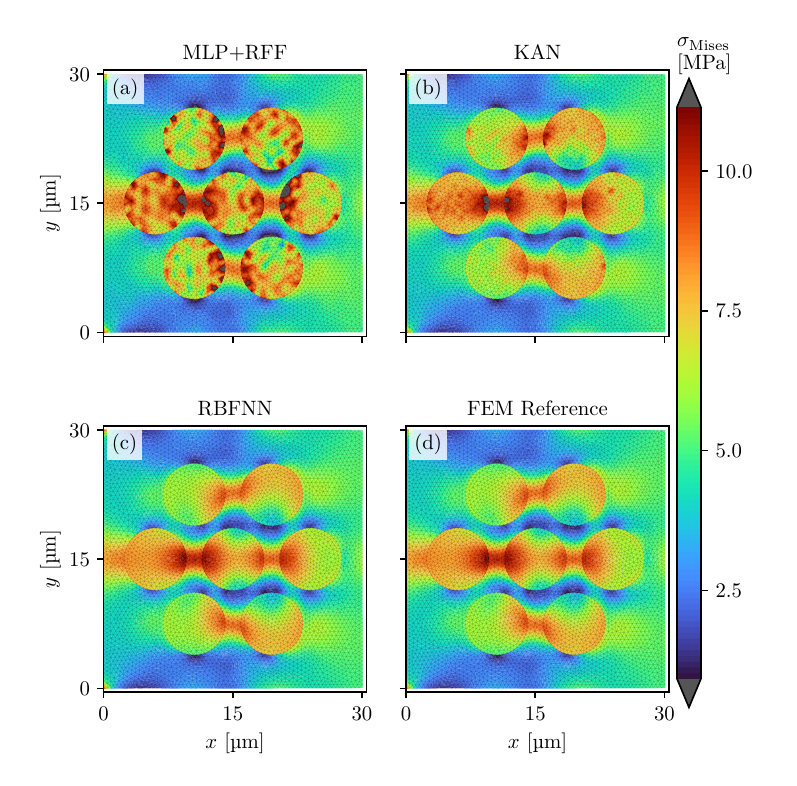}
    \caption{Shear stress ($\sigma_{xy}$) solutions for the \gls{cfrp}: (a-c) the \gls{mnem} with different neural network architectures and (d) the \gls{fem} reference solution.}
    \label{fig:fiber_sigma}
\end{figure}

Fig.~\ref{fig:fiber_sigma} illustrates the von-Mises stress distribution for the \gls{cfrp}. In contrast to the previous case studies, where the \gls{mlp}+\gls{rff} and \gls{kan} provided reasonable approximations, the high absolute stiffness values and multiple inclusions in this example reveal significant stability limitations. The \gls{mlp}+\gls{rff} solution fails to capture the smooth stress distribution within the fibers, exhibiting high-frequency oscillatory artifacts. Similarly, while the \gls{kan} reduces the noise in the material matrix, it struggles to resolve the sharp discontinuities at the material interfaces, leading to patchy stress predictions within the inclusions. Conversely, the \gls{rbfnn} produces a smooth stress field that correctly captures the concentrations in the matrix bridges between fibers and demonstrates excellent agreement with the \gls{fem} reference solution.

\begin{figure}[!ht]
    \centering
    \includegraphics{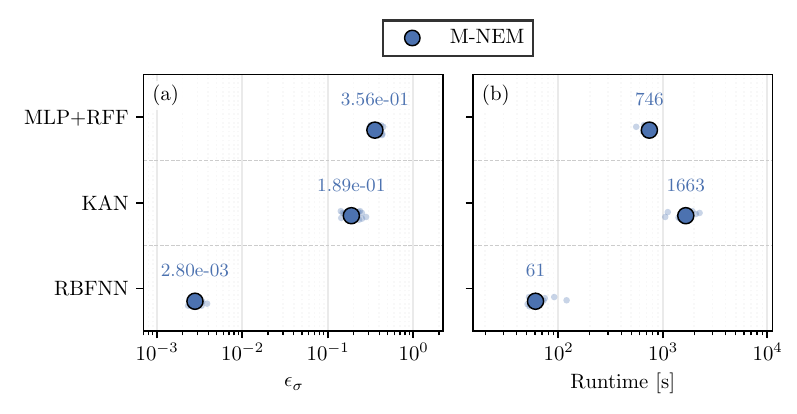}
    \caption{Comparison of relative  stress error (a) and runtime (b) for the \gls{mnem} with different neural network architectures on the \gls{cfrp}. Large dots show the median of the 30 training runs.}
    \label{fig:fiber_comparison}
\end{figure}

The quantitative performance is compared in Fig.~\ref{fig:fiber_comparison}. The stress error analysis confirms the qualitative observations, with the \gls{rbfnn} outperforming the \gls{kan} by almost two orders of magnitude. Notably, the runtime analysis demonstrates the computational efficiency of the \gls{rbfnn} architecture, which achieves a high-fidelity solution more than ten times faster than the \gls{kan}. This performance gap stems from the architectural nature of the models. The multi-layer composition of deep neural networks leads to a non-convex optimization landscape, whereas the fixed-center \gls{rbfnn} yields a displacement field that is linear with respect to the trainable weights. 

\begin{figure}[!ht]
    \centering
    \includegraphics{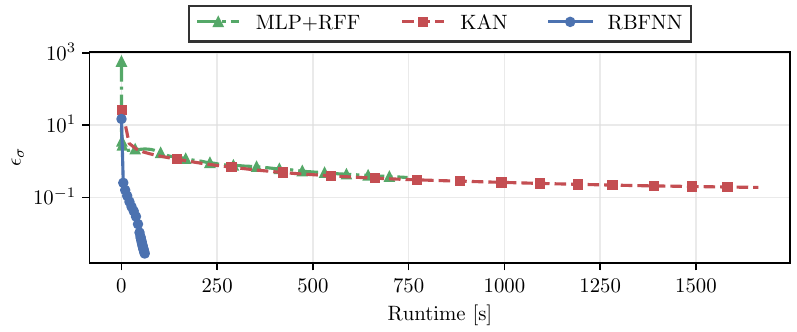}
    \caption{Evolution of the relative  stress error over runtime for the \gls{cfrp}.}
    \label{fig:fiber_evolution}
\end{figure}

This theoretical advantage is empirically demonstrated in Fig.~\ref{fig:fiber_evolution}, which plots the evolution of the global stress error $\epsilon_{\B\sigma}$ against runtime. The convergence behavior reveals a fundamental difference in the optimization process. The \gls{rbfnn} exhibits a rapid drop in error, characteristic of a well-conditioned problem, reaching high-fidelity solutions ($\epsilon_{\B\sigma} < 10^{-2}$) within seconds. Conversely, the \gls{kan} and \gls{mlp}+\gls{rff} display the slow, shallow decay typical of ill-conditioned systems, where the optimizer traverses a flat loss landscape. This confirms that, in addition to parallel efficiency, the speedup is driven by superior optimization dynamics.

\begin{figure}[!ht]
    \centering
    \includegraphics{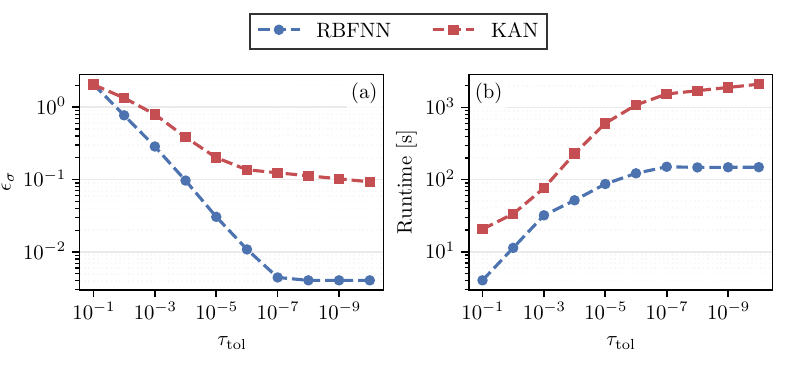}
    \caption{Comparison of relative  stress error (a) and runtime (b) for different convergence tolerances for the \gls{mnem} with \gls{kan} and \gls{rbfnn} on the \gls{cfrp}.}
    \label{fig:fiber_tolerance}
\end{figure}

Given the computational demands of this high-stiffness benchmark, the trade-off between accuracy and computational effort is further examined in Fig.~\ref{fig:fiber_tolerance}. The results indicate that computational costs can be substantially reduced by relaxing the convergence tolerance $\tau_{tol}$. This is especially relevant for the \gls{kan}, which carries a higher per-iteration cost. Conversely, the \gls{rbfnn} maintains high efficiency even under strict convergence criteria, benefiting from the parallelization potential of the single-layer architecture.

\section{Conclusions and Outlook}
\label{sec:conclusion}

This contribution proposed the \gls{mnem}, addressing fundamental limitations of the \gls{cnem} by introducing a mesh-based discretization. 
Primarily, it mitigated geometric overfitting, where expressive neural networks produce physically inadmissible displacement oscillations that remain invisible to the \gls{cnem} loss function. The \gls{mnem} suppressed these artifacts through shape function interpolation, which constrains the displacement field between nodes. Furthermore, this discretization enabled the replacement of automatic differentiation and Newton-Cotes integration with algebraic shape function derivatives and high-order Gaussian quadrature. This yielded higher accuracy, numerical stability, and, specifically for the \gls{rbfnn} architecture, a reduction in memory requirements.

The method was tested on heterogeneous microstructures, including a coupon with a stiff inclusion, a bioinspired calcified hydrogel unit cell, and a \gls{cfrp} microstructure. These benchmarks presented significant challenges to the \gls{cnem} due to high stiffness contrasts and complex nonlinear constitutive behavior. In the first case study, the \gls{mnem} greatly outperformed the standard \gls{cnem}, increasing solution accuracy by up to three orders of magnitude while decreasing computational effort by up to two orders of magnitude. This efficiency was attributed to both improved training stability, requiring fewer epochs for convergence, and a reduced computational cost per epoch. In the remaining two case studies, the standard \gls{cnem} failed to converge entirely, whereas the \gls{mnem} maintained stable and robust convergence. 

Furthermore, this study provided a comparative analysis of neural architectures for physics-informed elastostatics. While standard \glspl{mlp} suffer from spectral bias, the results indicate that this limitation can be overcome through informed architecture selection. Specifically, the \gls{rbfnn} emerged as the superior architecture, achieving the highest accuracy across all benchmarks while maintaining low computational cost. By employing fixed centers, the \gls{rbfnn} yielded a well-conditioned optimization landscape that avoided the local minima common to deep \glspl{kan} and \glspl{mlp} in high-stiffness regimes. The \gls{mnem} was crucial in enabling this architecture, as it removed the memory bottlenecks associated with differentiating through the kernel layer, allowing the network to scale to the densities required for high-fidelity simulation.
 
Beyond pure accuracy, the \gls{mnem} demonstrated a high degree of numerical robustness. Unlike its meshfree counterpart, the framework converged reliably with coarse discretizations and across a wide range of hyperparameters where the standard \gls{cnem} failed. This inherent stability reduced the need for extensive hyperparameter tuning, making the method more accessible for practical applications.
 
Despite these advancements, the computational cost of the \gls{mnem} for forward problems remains higher than that of optimized legacy \gls{fem} solvers. Future work will focus on extending the framework to 3D problems and combining it with neural mesh optimization algorithms such as the HiDeNN-FEM. However, the potential of the \gls{mnem} extends well beyond forward analysis. With its fully differentiable framework, the method establishes a robust foundation for inverse problems, material characterization, and the straightforward implementation of energy-based constitutive models. In these domains, where traditional methods often face computational or implementation bottlenecks, the proposed framework holds promise as a flexible and differentiable alternative.

\section*{CRediT authorship contribution statement}
\textbf{Pius J. M. Wichmann}: Conceptualization, Methodology, Software, Validation, Formal analysis, Investigation, Data Curation, Writing -- original draft, Visualization. 
\textbf{Stefan Hildebrand}: Conceptualization, Methodology, Validation, Formal analysis, Writing -- review \& editing, Supervision. 
\textbf{Sandra Klinge}: Conceptualization, Resources, Writing -- review \& editing, Supervision, Project administration, Funding acquisition.

\section*{Declaration of competing interest}
The authors declare that they have no known competing financial interests or personal relationships that could have appeared to influence the work reported in this paper.

\section*{Acknowledgment}
This work was supported by the German Academic Exchange Service (DAAD) [grant number 57762379].

\section*{Data availability}
The data used in this study is generated procedurally using the source code, which will be made available in a public repository upon acceptance.

\bibliographystyle{elsarticle-num}
\bibliography{bibliography}

\end{document}

%% file: tikz_nem_flowchart.tex
\begin{tikzpicture}[
    >=stealth, 
    box/.style={rectangle, rounded corners, draw=black!70, thick, minimum height=14mm, minimum width=22mm, align=center, fill=gray!5, inner sep=4pt},
    diff box/.style={rectangle, rounded corners, draw=black!90, thick, minimum height=14mm, minimum width=22mm, align=center, fill=gray!30, inner sep=4pt},
    loss box/.style={rectangle, rounded corners, draw=black!90, thick, minimum height=14mm, minimum width=11.5cm, align=center, fill=gray!30},
    arrow/.style={->, thick, draw=black!70},
]

    \def\xNN{-1.2}
    \def\xDist{1.8}
    \def\xAD{4.8}
    \def\xMat{7.8}
    \def\xLoss{3.15} 

    \node (X) at (-3.4, 0) {$\B X$};
    
    \node[box]      (NN)   at (\xNN, 0)   {Neural\\network};
    \node[box]      (Dist) at (\xDist, 0) {Dirichlet\\BCs};
    \node[diff box] (AD)   at (\xAD, 0)   {Derivative};
    \node[box]      (Mat)  at (\xMat, 0)  {Constitutive\\model};

    \node[loss box] (Loss) at (\xLoss, -3.2) {Integration (Total potential energy)};

    \draw[arrow] (X)  -- (NN);
    
    \draw[arrow] (NN)  -- node[midway, above] {$\hat{\B z}$} (Dist);
    \draw[arrow] (Dist) -- node[midway, above] {$\hat{\B u}$} coordinate[pos=0.45] (m1) (AD);
    \draw[arrow] (AD) -- node[midway, above] {$\hat{\B F}$} (Mat);

    \def\labelY{-1.4} 

    \draw[arrow] (m1) -- (m1 |- Loss.north);
    \node[fill=white, align=center, inner sep=1.5pt, font=\small] at (m1 |- 0, \labelY) {External work\\(Neumann BCs)};
    
    \draw[arrow] (Mat.south) -- (Mat.south |- Loss.north);
    \node[fill=white, align=center, inner sep=1.5pt, font=\small] at (Mat.south |- 0, \labelY) {Strain\\energy $\psi$};

    \draw[arrow, dashed, rounded corners=6pt] 
        (Loss.west) -- ++(-0.6, 0) coordinate (opt1)
        |- (\xNN, -2.0) coordinate (opt2)
        -- (NN.south);
        
    \path (opt1 |- opt2) -- (opt2) node[midway, above=1pt, align=center, font=\small] {Optimizer\\[-0.5ex]updates $\B \theta$};

\end{tikzpicture}

%% file: tikz_coupon.tex
\begin{tikzpicture}[
    >=stealth,
    matrix body/.style={draw=black, thick, fill=gray!10},
    inclusion body/.style={draw=black, thick, fill=gray!35},
    clamped line/.style={draw=black, very thick},
    traction/.style={->, thick, draw=black},
    dim/.style={<->, thin, draw=black!70},
    dim ext/.style={thin, draw=black!40}
]
    \def\W{4.6} 
    \def\H{2.3} 
    
    \pgfmathsetmacro{\IncXOne}{0.375*\W}
    \pgfmathsetmacro{\IncXTwo}{0.625*\W}
    \pgfmathsetmacro{\IncYOne}{0.25*\H}
    \pgfmathsetmacro{\IncYTwo}{0.75*\H}

    \pgfdeclarelayer{background}
    \pgfsetlayers{background,main}

    \begin{scope}
        \begin{scope}
            \clip (-0.25, 0) rectangle (0, \H);
            \foreach \i in {0, ..., 15} {
                \pgfmathsetmacro{\y}{\i * 0.15}
                \draw[black!70, thin] (0, \y) -- (-0.3, \y - 0.3);
            }
        \end{scope}
        \draw[clamped line] (0, 0) -- (0, \H);

        \draw[matrix body] (0, 0) rectangle (\W, \H);

        \draw[inclusion body] (\IncXOne, \IncYOne) rectangle (\IncXTwo, \IncYTwo);

        \node[right] at (0.1, \H - 0.35) {$E_1, \nu$};
        \node[align=center, inner sep=0pt] at (\W/2, \H/2) {$E_2, \nu$};

        \begin{scope}[shift={(-1.5, -0.5)}]
            \draw[->, thick, draw=black!80] (0,0) -- (0.6, 0) node[right, text=black] {$\mathrm{x}$};
            \draw[->, thick, draw=black!80] (0,0) -- (0, 0.6) node[above, text=black] {$\mathrm{y}$};
        \end{scope}

        \foreach \i in {0,1,2,3,4,5,6} {
            \draw[traction] (\W, \H*\i/6) -- (\W + 0.5, \H*\i/6);
        }
        \node[right, rotate=90] at (\W + 0.8, \H/4) {$\bar{t}=0.5$};
        
        \begin{pgfonlayer}{background}
            \draw[dim ext] (0, \H) -- (0, \H + 0.5);
            \draw[dim ext] (\W, \H) -- (\W, \H + 0.5);
            \draw[dim] (0, \H + 0.25) -- (\W, \H + 0.25) node[midway, fill=white, inner sep=2pt] {$L=2$};
            \draw[dim ext] (0, 0) -- (-0.9, 0);
            \draw[dim ext] (0, \H) -- (-0.9, \H);
            \draw[dim] (-0.75, 0) -- (-0.75, \H) node[midway, fill=white, inner sep=2pt, rotate=90] {$H=1$};
        \end{pgfonlayer}
    \end{scope}

\end{tikzpicture}

%% file: tikz_hydrogel.tex
\begin{tikzpicture}[
    >=stealth,
    matrix body/.style={draw=black, thick, fill=gray!10},
    calcium body/.style={draw=black, thick, fill=gray!40},
    pore body/.style={draw=black, thick, fill=white},
    clamped line/.style={draw=black, very thick},
    traction/.style={->, thick, draw=black},
    dim/.style={<->, thin, draw=black!70},
    dim ext/.style={thin, draw=black!40}
]
    \def\W{4.0}
    \def\H{4.0}
    
    \def\Rcalcium{0.25}
    \def\Rpore{0.18}

    \pgfdeclarelayer{background}
    \pgfsetlayers{background,main}

    \begin{scope}[shift={(0,0)}]
        \begin{scope}
            \clip (-0.25, 0) rectangle (0, \H);
            \foreach \i in {0, ..., 27} {
                \pgfmathsetmacro{\y}{\i * 0.15}
                \draw[black!70, thin] (0, \y) -- (-0.3, \y - 0.3);
            }
        \end{scope}
        \draw[clamped line] (0, 0) -- (0, \H);

        \draw[matrix body] (0, 0) rectangle (\W, \H);
        
        \draw[calcium body] (\W/2, \H/2) circle ({\Rcalcium*\W});
        
        \draw[pore body] (\W/2, \H/2) circle ({\Rpore*\W});

        \node[, inner sep=1pt] (LHydro) at (0.9, \H - 0.5) {Hydrogel};
        \draw[->, thin, draw=black!80] (LHydro.south) -- (1.2, \H - 1.2);
        
        \node[, inner sep=1pt] (LCalc) at (\W - 0.8, \H - 0.5) {Calcium};
        \draw[->, thin, draw=black!80] (LCalc.south) -- (\W/2 + 0.6, \H/2 + 0.6);
        
        \node[, inner sep=1pt] (LPore) at (\W - 0.8, 0.4) {Pore};
        \draw[->, thin, draw=black!80] (LPore.north) -- (\W/2 + 0.15, \H/2 - 0.15);

        \begin{scope}[shift={(-1.5, -0.5)}]
            \draw[->, thick, draw=black!80] (0,0) -- (0.6, 0) node[right, text=black] {$\mathrm{x}$};
            \draw[->, thick, draw=black!80] (0,0) -- (0, 0.6) node[above, text=black] {$\mathrm{y}$};
        \end{scope}

        \foreach \i in {0,1,2,3,4,5,6} {
            \draw[traction] (\W, \H*\i/6) -- (\W + 0.5, \H*\i/6);
        }
        \node[right, rotate=90] at (\W + 0.8, \H/4) {$\bar{t}=5\:\mathrm{kPa}$};

        \begin{pgfonlayer}{background}
            \draw[dim ext] (0, \H) -- (0, \H + 0.5);
            \draw[dim ext] (\W, \H) -- (\W, \H + 0.5);
            \draw[dim] (0, \H + 0.25) -- (\W, \H + 0.25) node[midway, fill=white, inner sep=2pt] {$L = 100\,\upmu\text{m}$};

            \draw[dim ext] (0, 0) -- (-0.9, 0);
            \draw[dim ext] (0, \H) -- (-0.9, \H);
            \draw[dim] (-0.75, 0) -- (-0.75, \H) node[midway, fill=white, inner sep=2pt, rotate=90] {$H = 100\,\upmu\text{m}$};
        \end{pgfonlayer}
    \end{scope}

\end{tikzpicture}

%% file: tikz_fiber.tex
\begin{tikzpicture}[
    >=stealth,
    matrix body/.style={draw=black, thick, fill=gray!10},
    fiber/.style={draw=black, thick, fill=gray!40},
    clamped line/.style={draw=black, very thick},
    traction/.style={->, thick, draw=black},
    dim/.style={<->, thin, draw=black!70},
    dim ext/.style={thin, draw=black!40}
]

    \def\W{4.0}
    \def\H{4.0}

    \newcommand{\drawnormfiber}[3]{
        \draw[fiber] ({(#1)*\W}, {(#2)*\H}) circle ({(#3)*\W});
    }

    \pgfdeclarelayer{background}
    \pgfsetlayers{background,main}

    \begin{scope}[shift={(0,0)}]
        \begin{scope}
            \clip (-0.25, 0) rectangle (0, \H);
            \foreach \i in {0, ..., 27} {
                \pgfmathsetmacro{\y}{\i * 0.15}
                \draw[black!70, thin] (0, \y) -- (-0.3, \y - 0.3);
            }
        \end{scope}
        \draw[clamped line] (0, 0) -- (0, \H);

        \draw[matrix body] (0, 0) rectangle (\W, \H);
        \begin{scope}
            \clip (0,0) rectangle (\W, \H);
            \foreach \x/\y in {0.50/0.50, 0.20/0.50, 0.80/0.50, 0.35/0.25, 0.65/0.25, 0.35/0.75, 0.65/0.75} {
                \drawnormfiber{\x}{\y}{0.12}
            }
        \end{scope}
        \draw[draw=black, thick] (0,0) rectangle (\W, \H);

        \draw[dim] ({0.35*\W - 0.12*\W}, {0.75*\H}) 
                -- ({0.35*\W + 0.12*\W}, {0.75*\H}) 
                   node[midway, fill=gray!40, inner sep=1.5pt] {$D$};

        \node[inner sep=1pt] at (0.8, 0.4) {Matrix};
        \draw[->, thin, draw=black!80] (0.8, 0.55) -- (0.5, 1.2);
        \node[inner sep=1pt] at (\W-0.8, \H-0.3) {Fibers};
        \draw[->, thin, draw=black!80] (\W-0.8, \H-0.45) -- (\W-1.2, \H-1.1);

        \begin{scope}[shift={(-1.5, -0.5)}]
            \draw[->, thick, draw=black!80] (0,0) -- (0.6, 0) node[right, text=black] {$\mathrm{x}$};
            \draw[->, thick, draw=black!80] (0,0) -- (0, 0.6) node[above, text=black] {$\mathrm{y}$};
        \end{scope}

        \foreach \y in {0.5, 1.2, 1.9, 2.6, 3.3, 4.0} {
            \draw[traction] (\W + 0.2, \y) -- (\W + 0.2, \y - 0.5) ;
        }
        \node[right, rotate=90] at (\W + 0.8, \H/4) {$\bar{t}=1\:\mathrm{MPa}$};

        \begin{pgfonlayer}{background}
            \draw[dim ext] (0, \H) -- (0, \H + 0.5);
            \draw[dim ext] (\W, \H) -- (\W, \H + 0.5);
            \draw[dim] (0, \H + 0.25) -- (\W, \H + 0.25) node[midway, fill=white, inner sep=2pt] {$L = 30\,\upmu\text{m}$};
            \draw[dim ext] (0, 0) -- (-0.9, 0);
            \draw[dim ext] (0, \H) -- (-0.9, \H);
            \draw[dim] (-0.75, 0) -- (-0.75, \H) node[midway, fill=white, inner sep=2pt, rotate=90] {$H = 30\,\upmu\text{m}$};
        \end{pgfonlayer}
    \end{scope}

\end{tikzpicture}

%% file: bibliography.bib
@article{karniadakis2021physics,
  title={Physics-informed machine learning},
  author={Karniadakis, George Em and Kevrekidis, Ioannis G and Lu, Lu and Perdikaris, Paris and Wang, Sifan and Yang, Liu},
  journal={Nature Reviews Physics},
  volume={3},
  number={6},
  pages={422--440},
  year={2021},
  doi={10.1038/s42254-021-00314-5}
}

@phdthesis{rauner2017,
  author={Jan Nicolas Rauner},
  title={Funktionale anorganisch/organische Hybridmaterialien f{\"u}r ultrasteife Hydrogele und antimikrobielle Lotusoberfl{\"a}chen},
  school={Technische Universit{\"a}t Dortmund},
  year={2017}
}

@article{ayguen2024,
  author={Ayg{\"u}n, Serhat and Klinge, Sandra},
  title={Two-scale computational homogenization of calcified hydrogels},
  journal={Mathematical Methods in the Applied Sciences},
  volume={47},
  number={17},
  pages={13182--13198},
  year={2024},
  doi={10.1002/mma.9095}
}

@article{yang2025,
  title={The influence of tensile rate on stress--strain test results of hydrogels},
  journal={Polymer Bulletin},
  volume={82},
  pages={8169--8179},
  year={2025},
  doi={10.1007/s00289-025-05873-6},
  author={Xia Yang and Xiaoli Ku and Zhihan Liu and Xu Xiang}
}

@article{ogden2004,
  author={Ogden, R. W. and Saccomandi, G. and Sgura, I.},
  title={Fitting hyperelastic models to experimental data},
  journal={Computational Mechanics},
  volume={34},
  pages={484--502},
  year={2004},
  doi={10.1007/s00466-004-0593-y}
}

@article{nguyenthanh2020,
  title={A deep energy method for finite deformation hyperelasticity},
  journal={European Journal of Mechanics - A/Solids},
  volume={80},
  pages={103874},
  year={2020},
  doi={10.1016/j.euromechsol.2019.103874},
  author={Vien Minh Nguyen-Thanh and Xiaoying Zhuang and Timon Rabczuk}
}

@article{fuhg2022,
  title={The mixed Deep Energy Method for resolving concentration features in finite strain hyperelasticity},
  journal={Journal of Computational Physics},
  volume={451},
  pages={110839},
  year={2022},
  doi={10.1016/j.jcp.2021.110839},
  author={Jan N. Fuhg and Nikolaos Bouklas}
}

@misc{dolfinx,
  title={{DOLFINx}: the next generation {FEniCS} problem solving environment},
  author={Baratta, Igor A. and Dean, Joseph P. and Dokken, J{\o}rgen S. and Habera, Michal and Hale, Jack S. and Richardson, Chris N. and Rognes, Marie E. and Scroggs, Matthew W. and Sime, Nathan and Wells, Garth N.},
  doi={10.5281/zenodo.10447666},
  year={2023}
}

@article{basix1,
  title={Construction of arbitrary order finite element degree-of-freedom maps on polygonal and polyhedral cell meshes},
  author={Scroggs, Matthew W. and Dokken, J{\o}rgen S. and Richardson, Chris N. and Wells, Garth N.},
  journal={ACM Transactions on Mathematical Software},
  year={2022},
  volume={48},
  number={2},
  pages={18:1--18:23},
  doi={10.1145/3524456}
}

@article{basix2,
  title={Basix: a runtime finite element basis evaluation library},
  author={Scroggs, Matthew W. and Baratta, Igor A. and Richardson, Chris N. and Wells, Garth N.},
  journal={Journal of Open Source Software},
  year={2022},
  volume={7},
  number={73},
  pages={3982},
  doi={10.21105/joss.03982}
}

@article{ufl,
  title={Unified Form Language: A domain-specific language for weak formulations of partial differential equations},
  author={Alnaes, Martin S. and Logg, Anders and {\O}lgaard, Kristian B. and Rognes, Marie E. and Wells, Garth N.},
  journal={{ACM} Transactions on Mathematical Software},
  year={2014},
  volume={40},
  doi={10.1145/2566630}
}

@article{chadha2023,
  author={Charul Chadha and Junyan He and Diab Abueidda and Seid Koric and Erman Guleryuz and Iwona Jasiuk},
  title={Improving the accuracy of the deep energy method},
  journal={Acta Mechanica},
  year={2023},
  volume={234},
  pages={5975--5998},
  doi={10.1007/s00707-023-03691-3}
}

@article{he2023elastoplasticity,
  title={A deep learning energy-based method for classical elastoplasticity},
  volume={162},
  doi={10.1016/j.ijplas.2023.103531},
  journal={International Journal of Plasticity},
  author={He, Junyan and Abueidda, Diab and Abu Al-Rub, Rashid and Koric, Seid and Jasiuk, Iwona},
  year={2023},
  pages={103531}
}

@article{heGNN2023,
  author={He, Junyan and Abueidda, Diab and Koric, Seid and Jasiuk, Iwona},
  title={On the use of graph neural networks and shape-function-based gradient computation in the deep energy method},
  journal={International Journal for Numerical Methods in Engineering},
  volume={124},
  number={4},
  pages={864--879},
  doi={10.1002/nme.7146},
  year={2023}
}

@article{wang2025kan,
  title={Kolmogorov--Arnold-Informed neural network: A physics-informed deep learning framework for solving forward and inverse problems based on Kolmogorov--Arnold Networks},
  journal={Computer Methods in Applied Mechanics and Engineering},
  volume={433},
  pages={117518},
  year={2025},
  doi={10.1016/j.cma.2024.117518},
  author={Yizheng Wang and Jia Sun and Jinshuai Bai and Cosmin Anitescu and Mohammad Sadegh Eshaghi and Xiaoying Zhuang and Timon Rabczuk and Yinghua Liu}
}

@article{wang2021,
  title={On the eigenvector bias of Fourier feature networks: From regression to solving multi-scale PDEs with physics-informed neural networks},
  journal={Computer Methods in Applied Mechanics and Engineering},
  volume={384},
  pages={113938},
  year={2021},
  doi={10.1016/j.cma.2021.113938},
  author={Sifan Wang and Hanwen Wang and Paris Perdikaris}
}

@misc{liu2025,
  title={{KAN}: Kolmogorov--Arnold Networks},
  author={Ziming Liu and Yixuan Wang and Sachin Vaidya and Fabian Ruehle and James Halverson and Marin Solja{\v{c}}i{\'c} and Thomas Y. Hou and Max Tegmark},
  year={2025},
  doi={10.48550/arXiv.2404.197566}
}

@misc{kingma2017,
  title={Adam: A Method for Stochastic Optimization},
  author={Diederik P. Kingma and Jimmy Ba},
  year={2017},
  doi={10.48550/arXiv.1412.6980}
}

@article{liu1989,
  title={On the limited memory {BFGS} method for large scale optimization},
  author={Liu, Dong C. and Nocedal, Jorge},
  year={1989},
  doi={10.1007/BF01589116},
  volume={45},
  pages={503--528},
  journal={Mathematical Programming},
  number={1}
}

@inproceedings{tancik2020,
  author={Tancik, Matthew and Srinivasan, Pratul P. and Mildenhall, Ben and Fridovich-Keil, Sara and Raghavan, Nithin and Singhal, Utkarsh and Ramamoorthi, Ravi and Barron, Jonathan T. and Ng, Ren},
  title={Fourier features let networks learn high frequency functions in low dimensional domains},
  year={2020},
  booktitle={Proceedings of the 34th International Conference on Neural Information Processing Systems},
  pages={7537--7547}
}

@article{zhao2025,
  title={Physically informed neural networks for homogenization and localization of composites with periodic microstructures},
  journal={Composite Structures},
  volume={367},
  pages={119260},
  year={2025},
  doi={10.1016/j.compstruct.2025.119260},
  author={Wenhui Zhao and Ruixuan Hao and Ming Zhang and Qiang Chen and Zhibo Yang and Xuefeng Chen}
}

@article{liang2022,
  title={{SEM}: a shallow energy method for finite deformation hyperelasticity problems},
  journal={Acta Mechanica},
  volume={233},
  pages={1739--1755},
  year={2022},
  doi={10.1007/s00707-022-03174-x},
  author={Liang, Z. and Gao, H. and Li, T.}
}

@misc{kharazmi2019,
  title={Variational Physics-Informed Neural Networks For Solving Partial Differential Equations},
  author={Ehsan Kharazmi and Z. Zhang and George Em Karniadakis},
  year={2019},
  doi={10.48550/arXiv.1912.00873}
}

@article{raissi2019,
  title={Physics-informed neural networks: A deep learning framework for solving forward and inverse problems involving nonlinear partial differential equations},
  journal={Journal of Computational Physics},
  volume={378},
  pages={686--707},
  year={2019},
  doi={10.1016/j.jcp.2018.10.045},
  author={M. Raissi and P. Perdikaris and George Em Karniadakis}
}

@article{samaniego2020,
  title={An energy approach to the solution of partial differential equations in computational mechanics via machine learning: Concepts, implementation and applications},
  journal={Computer Methods in Applied Mechanics and Engineering},
  volume={362},
  pages={112790},
  year={2020},
  doi={10.1016/j.cma.2019.112790},
  author={E. Samaniego and C. Anitescu and S. Goswami and V.M. Nguyen-Thanh and H. Guo and K. Hamdia and X. Zhuang and T. Rabczuk}
}

@article{graham2024,
  title={Multiscale homogenisation of diffusion in enzymatically-calcified hydrogels},
  journal={Journal of the Mechanical Behavior of Biomedical Materials},
  volume={149},
  pages={106244},
  year={2024},
  doi={10.1016/j.jmbbm.2023.106244},
  author={Marc Graham and Sandra Klinge}
}

@phdthesis{burchak2021,
  author={Vadym Burchak},
  title={Charakterisierung der Hydrogele {GelNB/S} und {GelNB/DTT} f{\"u}r das {3D} Bioprinting von Knochen-, Knorpel- und Fettersatzgeweben},
  school={Albert-Ludwigs-Universit{\"a}t Freiburg},
  year={2021},
  doi={10.6094/UNIFR/223204}
}

@article{anandh2025,
  author={Anandh, Thivin and Ghose, Divij and Jain, Himanshu and Ganesan, Sashikumaar},
  title={{FastVPINNs}: Tensor-Driven Acceleration of {VPINNs} for Complex Geometries},
  journal={SIAM Journal on Scientific Computing},
  volume={47},
  number={3},
  pages={C578--C600},
  year={2025},
  doi={10.1137/24M1658620}
}

@article{radin2023,
  author={Radin, Nils and Klinkel, Sven and Altay, Okyay},
  title={Effects of variational formulations on physics-informed neural network performance in solid mechanics},
  journal={PAMM},
  volume={23},
  number={4},
  pages={e202300222},
  doi={10.1002/pamm.202300222},
  year={2023}
}

@article{berronne2022,
  title={Variational {Physics} {Informed} {Neural} {Networks}: the {Role} of {Quadratures} and {Test} {Functions}},
  volume={92},
  doi={10.1007/s10915-022-01950-4},
  number={3},
  journal={Journal of Scientific Computing},
  author={Berrone, Stefano and Canuto, Claudio and Pintore, Moreno},
  year={2022},
  pages={100}
}

@article{brumand2025,
  title={Advancing deformation calculation: a physics-informed deep graph learning framework for hyperelastic materials},
  volume={12},
  doi={10.1186/s40323-025-00304-7},
  number={1},
  journal={Advanced Modeling and Simulation in Engineering Sciences},
  author={Brumand-Poor, Faras and Trautmann, Matthias and Schmitz, Katharina},
  year={2025},
  pages={20}
}

@article{gao2022,
  title={Physics-informed graph neural {Galerkin} networks: A unified framework for solving {PDE}-governed forward and inverse problems},
  journal={Computer Methods in Applied Mechanics and Engineering},
  volume={390},
  pages={114502},
  year={2022},
  doi={10.1016/j.cma.2021.114502},
  author={Han Gao and Matthew J. Zahr and Jian-Xun Wang}
}

@article{herrmann2024,
  title={Deep learning in computational mechanics: a review},
  volume={74},
  doi={10.1007/s00466-023-02434-4},
  number={2},
  journal={Computational Mechanics},
  author={Herrmann, Leon and Kollmannsberger, Stefan},
  year={2024},
  pages={281--331}
}

@book{kollmannsberger2025,
  author={Herrmann, Leon and Jokeit, Moritz and Weeger, Oliver and Kollmannsberger, Stefan},
  title={Deep Learning in Computational Mechanics: An Introductory Course},
  publisher={Springer},
  address={Cham},
  year={2025},
  edition={2},
  doi={10.1007/978-3-031-89529-6}
}

@misc{wang2025xdem,
  title={Towards Unified {AI}-Driven Fracture Mechanics: The Extended Deep Energy Method ({XDEM})},
  author={Yizheng Wang and Yuzhou Lin and Somdatta Goswami and Luyang Zhao and Huadong Zhang and Jinshuai Bai and Cosmin Anitescu and Mohammad Sadegh Eshaghi and Xiaoying Zhuang and Timon Rabczuk and Yinghua Liu},
  year={2025},
  doi={10.48550/arXiv.2511.05888}
}

@article{sukumar2022,
  title={Exact imposition of boundary conditions with distance functions in physics-informed deep neural networks},
  author={Sukumar, N. and Srivastava, Ankit},
  journal={Computer Methods in Applied Mechanics and Engineering},
  volume={389},
  pages={114333},
  year={2022},
  doi={10.1016/j.cma.2021.114333}
}

@article{nguyentanh2021,
  title={Parametric deep energy approach for elasticity accounting for strain gradient effects},
  journal={Computer Methods in Applied Mechanics and Engineering},
  volume={386},
  pages={114096},
  year={2021},
  doi={10.1016/j.cma.2021.114096},
  author={Vien Minh Nguyen-Thanh and Cosmin Anitescu and Naif Alajlan and Timon Rabczuk and Xiaoying Zhuang}
}

@article{thombre2025,
  title={Energy-based methods for solving forward and inverse linear elasticity problems in {2D} structures},
  journal={Computers \& Structures},
  volume={316},
  pages={107899},
  year={2025},
  doi={10.1016/j.compstruc.2025.107899},
  author={Manish Thombre and Cosmin Anitescu and BVSS Bharadwaja and Yizheng Wang and Timon Rabczuk and Alankar Alankar}
}

@book{holzapfel2000,
  author={Holzapfel, Gerhard A.},
  title={Nonlinear Solid Mechanics: A Continuum Approach for Engineering},
  publisher={John Wiley \& Sons},
  address={Chichester},
  year={2000}
}

@book{wriggers2008,
  author={Wriggers, Peter},
  title={Nonlinear Finite Element Methods},
  publisher={Springer},
  address={Berlin, Heidelberg},
  year={2008},
  doi={10.1007/978-3-540-71001-1}
}

@misc{kern,
  title={{Technical Data Sheet: Polyamid PA 6.6}},
  author={{Kern GmbH}},
  year={2024},
  note={Available at \url{https://www.kern.de/de/technisches-datenblatt/polyamid-pa-66?n=1142_1}. Accessed: 2025-12-04}
}

@manual{zoltek2021,
  title={Zoltek {PX35} Commercial Carbon Fiber},
  author={{Zoltek Corporation}},
  year={2021},
  note={Available at \url{https://zoltek.com/wp-content/uploads/2021-PX35-Brochure-1.pdf}. Accessed: 2025-12-04}
}

@book{chawla2019,
  title={Composite Materials: Science and Engineering},
  author={Chawla, Krishan K.},
  edition={4},
  year={2019},
  publisher={Springer},
  address={Cham},
  doi={10.1007/978-3-030-28983-6}
}

@article{shirinbayan2024,
  title={Effects of hygrothermal aging on the physicochemical and mechanical properties of {3D}-printed {PA6}},
  author={Shirinbayan, Mohammadali and Benfriha, Khaled and Ahmadifar, Mohammad and Penavayre, Clara and Nouira, Samia and Fitoussi, Joseph},
  journal={The International Journal of Advanced Manufacturing Technology},
  volume={131},
  pages={4811--4823},
  year={2024},
  doi={10.1007/s00170-024-13323-2}
}

@article{hildebrand2025,
  title={Coupled {CANN-DEM} simulation in solid mechanics},
  journal={Machine Learning: Science and Technology},
  volume={6},
  pages={015038},
  year={2025},
  author={Stefan Hildebrand and Jonathan Georg Friedrich and Melika Mohammadkhah and Sandra Klinge},
  doi={10.1088/2632-2153/adaf74}
}

@article{he2023topology,
  title={Deep energy method in topology optimization applications},
  author={Junyan He and Charul Chadha and Shashank Kushwaha and Seid Koric and Diab Abueidda and Iwona Jasiuk},
  year={2023},
  doi={10.1007/s00707-022-03449-3},
  volume={234},
  pages={1365--1379},
  journal={Acta Mechanica},
  number={4}
}

@article{goswami2020,
  title={Adaptive fourth-order phase field analysis using deep energy minimization},
  journal={Theoretical and Applied Fracture Mechanics},
  volume={107},
  pages={102527},
  year={2020},
  doi={10.1016/j.tafmec.2020.102527},
  author={Somdatta Goswami and Cosmin Anitescu and Timon Rabczuk}
}

@misc{paszke2019,
  title={{PyTorch}: An Imperative Style, High-Performance Deep Learning Library},
  author={Adam Paszke and Sam Gross and Francisco Massa and Adam Lerer and James Bradbury and Gregory Chanan and Trevor Killeen and Zeming Lin and Natalia Gimelshein and Luca Antiga and Alban Desmaison and Andreas K{\"o}pf and Edward Yang and Zach DeVito and Martin Raison and Alykhan Tejani and Sasank Chilamkurthy and Benoit Steiner and Lu Fang and Junjie Bai and Soumith Chintala},
  year={2019},
  doi={10.48550/arXiv.1912.01703}
}

@article{klinge2012,
  author={Klinge, Sandra and Hackl, Klaus},
  year={2012},
  title={Application of the multiscale {FEM} to the modeling of nonlinear composites with a random microstructure},
  volume={10},
  journal={International Journal for Multiscale Computational Engineering},
  doi={10.1615/IntJMultCompEng.2012002059},
  pages={213--227}
}

@article{sanjalawe2026,
	author = {Sanjalawe, Yousef and Al-E'mari, Salam and Makhadmeh, Sharif Naser},
	date = {2026/08/06},
	doi = {10.1007/s11831-026-10747-9},
	isbn = {1886-1784},
	journal = {Archives of Computational Methods in Engineering},
	title = {Physics-Informed Neural Networks in Computational Mechanics: A Critical Review of Stability, Generalization, and Multi-Scale Applications},
	year = {2026}
}

@article{zhang2021,
	author = {Zhang, Lei and Cheng, Lin and Li, Hengyang and Gao, Jiaying and Yu, Cheng and Domel, Reno and Yang, Yang and Tang, Shaoqiang and Liu, Wing Kam},
	date = {2021/01/01},
	doi = {10.1007/s00466-020-01928-9},
	isbn = {1432-0924},
	journal = {Computational Mechanics},
	number = {1},
	pages = {207--230},
	title = {Hierarchical deep-learning neural networks: finite elements and beyond},
	volume = {67},
	year = {2021}
}
